\documentclass[12pt,reqno,fleqn]{amsart}
\usepackage{latexsym}
\usepackage{amssymb}
\usepackage{amsxtra}
\usepackage{amsmath}
\usepackage{amsfonts}
\usepackage[dvips]{graphicx}
\usepackage{amsmath}
\usepackage{amssymb,amsmath,amsthm,amscd,mathrsfs,amsbsy,amsfonts}
\usepackage{pstricks,pst-node}
\usepackage{amsbsy,young,colortbl,cite}

\usepackage{graphicx}
\usepackage[metapost]{mfpic}
\usepackage{mflogo}

\usepackage{epstopdf}
\usepackage{graphics}
\usepackage{subfigure}

\def\@makecaption#1#2{\vskip\abovecaptionskip
  \sbox\@tempboxa{\small #1: #2}%
  \ifdim \wd\@tempboxa >\hsize \small #1: #2\par
  \else \global \@minipagefalse \hb@xt@\hsize{\hfil\box\@tempboxa\hfil}\fi
  \vskip\belowcaptionskip}
\newcommand{\cleqn}{\setcounter{equation}{0}}
\newcommand{\clth}{\setcounter{theorem}{0}}
\newcommand {\sectionnew}[1]{\section{#1}\cleqn\clth}
\makeatletter
    
    \newcommand{\Rmnum}[1]{\expandafter\@slowromancap\romannumeral #1@}
  \makeatother

\def\({\left(}
\def\){\right)}
\def\[{\begin{eqnarray}}
\def\]{\end{eqnarray}}

\begin{document}

\title{On Rogue wave in the Kundu-DNLS equation}
\author{Shibao Shan\dag, Chuanzhong Li\dag
, Jingsong He\dag\ddag}

\dedicatory {\small \dag Department of Mathematics,  Ningbo University, Ningbo, 315211, China\\
}
\thanks{\ddag Corresponding author}
\texttt{}

\date{}

%%%%%%%%%%%%%%%%%%%%%%%%%%%%%%%%%%%%%%%%%%%%%%%%
\begin{abstract}
In this paper, the determinant representation of the n-fold Darboux
transformation (DT) of the Kundu-DNLS equation is given. Based on
our analysis, the soliton solutions, positon solutions and breather
solutions of the Kundu-DNLS equation are given explicitly.
 Further, we also construct the  rogue wave solutions  which are given by using the Taylor expansion of the breather
 solution. Particularly, these rogue wave solutions possess several free parameters. With the help of these parameters, these rogue waves constitute several patterns, such as fundamental pattern,  triangular pattern, circular pattern.

%%%%%
\end{abstract}

%%%%%%%%%%%%%%%%%%%%%%%%%%%%%%%%%%%%%%%%%%%%%%%%
%%%%%%%%%%%%%%%%%%%%%%%%%%%%%%%%%%%%%%%%%%%%%%%%
\maketitle

\allowdisplaybreaks
 \setcounter{section}{0}

\sectionnew{Introduction }

It is well known that the derivative nonlinear Schr\"{o}dinger
(DNLS) equation
\begin{eqnarray}\label{sy1}
  iq_t+q_{xx}-i\alpha(|q|^2q)_x&=&0
\end{eqnarray}
is one of the most important integrable systems in many branches of
physics and applied mathematics, particularly in optics, water wave and so on\cite{NLSocean,SWX}. In the eq.(\ref{sy1}), $q$ represents complex field envelope and the subscripts imply the partial derivative with regard to $x$ or $t$. Considering the significance of the higher order nonlinearities in physical system,  the DNLS equation yields an integrable higher nonlinear equation, i.e. Kundu-DNLS equation\cite{KUNDU,KA}
\begin{eqnarray}\label{sy2}
iQ_t+Q_{xx}+i\alpha(Q^2Q^{\ast})_x-(\theta_t+\theta_x^2-i\theta_{xx})Q+\theta_x(2iQ_x-\alpha Q^2Q^{\ast})&=&0
\end{eqnarray}
by means of a nonlinear transformation of the field $q\rightarrow
Q=qe^{-i\theta}$ with arbitrary gauge function $\theta$.   Here $Q^{\ast}$ denotes the complex conjugate of Q, and $\alpha$ is a real parameter. For example, setting $\theta=\delta \int Q(x')^2dx'$,
Kundu-DNLS equation implies Eckhaus-Kundu (EK) equation\cite{KUNDU}. There are several method to solve the integrable equations, for instance, Hirota method\cite{hirota}, inverse scattering transformation \cite{AP}, bilinear method\cite{SK}, Darboux transformation\cite{matveev}. Among these methods, the Darboux transformation has an old history, which originates from the paper of Darboux in 1882 about the study of the Sturn-Liouville equation, and is a efficient method to construct the solutions for the integrable systems. To a given equation, there are also many approaches to find the corresponding Darboux transformation, such as the gauge transformation method\cite{matveev,li,he,HMB}, the operator factorization method\cite{LT}. In this paper, in terms of the Darboux transformation, we find that this equation can also be used to describe many intriguing physical phenomena, and possessing the soliton solutions, breather solution, positon solution, the rogue waves.

Before considering the Darboux transformation, let us briefly discuss the importance of rogue waves in mathematical field and physical field. Rogue waves, appearing in oceans ``appears from nowhere and disappears without a
trace\cite{akhmediev}", which cause a large number of disasters for people. Consequently, it appeals to much more and more attention for many scientists  and has been studied extensively in other fields such as optics\cite{tai1,tai2,solli}, Bose-Einstein condensates \cite{KB}, femtosecond pulse\cite{akhmediev4,nnakhmediev} propagation due to the modulation instability\cite{dudley}.
The higher order rogue waves of the nonlinear Schr\"{o}dinger (NLS) equation  have interesting patterns\cite{matveev2,akhmediev5,akhmediev6,landaudamping, HJS}. Recently, we have given the rogue waves of the DNLS and the coupled system  of  Hirota and Maxwell-Bloch equations\cite{SWX,rogueHMB}, Inspired by the importance of these recent interesting developments about the analysis of rogue waves of the NLS-type equations, we shall also construct the rogue wave solutions of the Kundu-DNLS equation with the help of the Darboux transformation.

The paper is organized as follows.  In Section 2, the Darboux transformation and the Lax
pair  of the Kundu-DNLS  equation will be introduced firstly.  We
also derived the one-fold Darboux transformation of the Kundu-DNLS  equation. And the
determinant-formed generalization of one-fold Darboux transformation to 2-fold Darboux transformation of the
Kundu-DNLS  equation will be given. In Section 3, by making use
of Darboux transformation, one soliton, two soliton,  positon solution, are derived. With the help of these formulas, breather solution, and
rogue waves are derived in Section 4. In Section 5, conclusions are given.

\sectionnew{Darboux transformation and lax pair }
The Darboux transformation is a powerful method used  to generate the soliton solutions for integrable equations. Inspired by classical Darboux transformation for the DNLS equation\cite{DNLStruncation,DNLS1,fan,KN,kawata1,huang4},  we consider the coupled Kundu-DNLS equation,
\begin{eqnarray}
iQ_t+Q_{xx}-i\alpha(Q^2R)_x-(\theta_t+\theta_x^2-i\theta_{xx})Q+\theta_x(2iQ_x+\alpha Q^2R)&=&0\label{sys11},
\end{eqnarray}
\begin{eqnarray}
iR_t-R_{xx}-i\alpha(R^2Q)_x+(\theta_t+\theta_x^2+i\theta_{xx})R+\theta_x(2iR_x-\alpha R^2Q)&=&0\label{sys12},
\end{eqnarray}
where $\theta$ is a arbitrary gauge function. This form of
the equation is very extensive, which is reduced to the eq.(\ref{sy2}) for  $R=-Q^{\ast}$ with the sign of the nonlinear term changed. The Kundu-DNLS equation can be obtained if $\alpha$ is a real parameter.

We first present a general framework for deriving the required conservation rule for the DNLS equation. We start with the linear set of Lax equations:
\begin{eqnarray}\label{ABC1}
\Phi_{x}&=&U\Phi,
\Phi_{t}=V\Phi,
\end{eqnarray}
where $U$ and $V$ depend on the complex constant eigenvalue parameter $\lambda$
\[
U&=&\-i\frac{\lambda^2}{4}\left(\begin{matrix}1& 0\\ 0& -1
\end{matrix}\right)+\frac{i}{2}\lambda\sqrt{\alpha}\left(\begin{matrix}0& Re^{-i\theta}\\ Qe^{i\theta}& 0
\end{matrix}\right),\\ \notag
V&=&i(\frac{\lambda^4}{8}-\frac{\alpha}{4}\lambda^2QR)\left(\begin{matrix}1& 0\\ 0& -1
\end{matrix}\right)+i\left(\begin{matrix}0& G^*\\ G& 0
\end{matrix}\right),\\ \notag
\]
with
$G=\frac{\lambda}{4}\sqrt{\alpha}(-\lambda^2Qe^{i\theta}+2i(Q_xe^{i\theta}+iQe^{i\theta}\theta_{x})+2\alpha Q^2Q^{\ast}e^{i\theta})$,
where $\lambda$ is the eigenvalue, $\Phi$ is the eigenfunction corresponding to $\lambda$.

Next, we will give the detailed proof for the one-fold Darboux transformation of Kundu-DNLS equation. Equations(\ref{sys11}) and (\ref{sys12}) are equivalent to the integrability condition  $U_{t}-V_{x}+[U,V]=0$ of (\ref{ABC1}). we would like to introduce a simple gauge transformation of the spectral problem (\ref{ABC1}) with the following form
\begin{eqnarray}\label{bh3}
\psi^{[1]}&=&T~\psi.
\end{eqnarray}

It can transform linear equation (\ref{ABC1}) into a new one possessing the same matrix form, namely,
\begin{eqnarray}\label{bh1}
{\psi^{[1]}}_{x}&=&U^{[1]}~\psi^{[1]},\ \ U^{[1]}=(T_{x}+T~U)T^{-1},
\end{eqnarray}
\begin{eqnarray}\label{bh2}
{\psi^{[1]}}_{t}&=&V^{[1]}~\psi^{[1]}, \ \ V^{[1]}=(T_{t}+T~V)T^{-1},
\end{eqnarray}\\
where $U^{[1]}$, $V^{[1]}$ have the similar forms as $U$, $V$. By cross differentiating (\ref{bh1}) and (\ref{bh2}), we obtain
\begin{eqnarray}\label{bh4}
{U^{[1]}}_{t}-{V^{[1]}}_{x}+[{U^{[1]}},{V^{[1]}}]&=&T(U_{t}-V_{x}+[U,V])T^{-1}.
\end{eqnarray}

\subsection{One-fold Darboux transformation}

In general, considering the universality of Darboux transformation, according to the Kundu-DNLS equation (\ref{sys11}) and (\ref{sys12}), we can start from
\begin{eqnarray}\label{tt1}
T&=&T(\lambda)=\left( \begin{array}{cc}
a_{2}&b_{2} \\
c_{2} &d_{2}\\
\end{array} \right)\lambda^{2}+\left( \begin{array}{cc}
a_{1}&b_{1} \\
c_{1} &d_{1}\\
\end{array} \right)\lambda+\left( \begin{array}{cc}
a_{0}&b_{0}\\
c_{0} &d_{0}\\
\end{array} \right),
\end{eqnarray}
where $a_{0},  b_{0},  c_{0},   d_{0},   a_{1},   b_{1},   c_{1},   d_{1},   a_{2},   b_{2},   c_{2},   d_{2}$ are functions of $x$, $t$. From \begin{eqnarray}\label{tt2} T_{x}+T~U=U^{[1]}~T,
\end{eqnarray}
comparing the coefficients of $\lambda^{j}, j=4, 3, 2, 1, 0$, it yields
\begin{eqnarray}\label{xx1}
&&\lambda^{4}: b_{2}=0,\ c_{2}=0,\\\nonumber
&&\lambda^{3}:-d_{2}\sqrt{\alpha}Q+c_{1}e^{i\theta}+a_{2}\sqrt{\alpha}Q^{[1]}=0, d_{2}\sqrt{\alpha}R^{[1]}-b_{1}e^{-i\theta}-a_{2}\sqrt{\alpha}R=0,\\\nonumber
&&\lambda^{2}:-d_{1}\sqrt{\alpha}Qe^{i\theta}+c_{0}+a_{1}\sqrt{\alpha}Q^{[1]}e^{i\theta}=0, d_{1}\sqrt{\alpha}R^{[1]}e^{-i\theta}-b_{0}-a_{1}\sqrt{\alpha}Re^{-i\theta}=0,\\\nonumber
&& \ \ \ \ \ ic_{1}\sqrt{\alpha}R^{[1]}e^{-i\theta}-2{a_{2}}_{x}-ib_{1}\sqrt{\alpha}Qe^{i\theta}=0, ib_{1}\sqrt{\alpha}Q^{[1]}e^{i\theta}-2{d_{2}}_{x}-ic_{1}\sqrt{\alpha}Re^{-i\theta}=0,\\\nonumber
&&\lambda^{1}:ic_{0}\sqrt{\alpha}R^{[1]}e^{-i\theta}-2a_{1x}-ib_{0}\sqrt{\alpha}Qe^{i\theta}=0, id_{0}\sqrt{\alpha}R^{[1]}e^{-i\theta}-2b_{1x}-ia_{0}\sqrt{\alpha}Re^{-i\theta}=0,\\\nonumber
&& \ \ \ \ \
ia_{0}\sqrt{\alpha}Q^{[1]}e^{i\theta}-2c_{1x}-id_{0}\sqrt{\alpha}Qe^{i\theta}=0, ib_{0}\sqrt{\alpha}Q^{[1]}e^{i\theta}-2d_{1x}-ic_{0}\sqrt{\alpha}Re^{-i\theta}=0,\\\nonumber
&&\lambda^{0}: {a_{0}}_{x}={b_{0}}_{x}={c_{0}}_{x}={d_{0}}_{x}=0.
\end{eqnarray}
The last equation shows $a_{0},b_{0},c_{0},d_{0} $ are functions of $t$ only.
Similarly, from
\begin{eqnarray}\label{tt3} T_{t}+T~V=V^{[1]}~T,
\end{eqnarray}
 comparing the coefficients of $\lambda^{j} ,j =4,3, 2, 1, 0$, in the same way, we can get
\begin{eqnarray}\label{ttt1}
&& b_{2}=0,\ c_{2}=0,{a_{0}}_{t}={b_{0}}_{t}={c_{0}}_{t}={d_{0}}_{t}=0.
\end{eqnarray}
The last equation shows $a_{0}, b_{0}, c_{0}, d_{0} $ are functions of  $x$ only. Therefore,
$a_{0}, b_{0}, c_{0}, d_{0} $ are constants.

In order to get the non-trivial solutions, we present a Darboux transformation under the condition $a_{1} = 0, d_{1} = 0, b_{0} = 0, c_{0} = 0$.
Based on eq.(\ref{xx1}) and eq.(\ref{ttt1}) and without losing any generality,
let  Darboux matrix $T$ be  in the  form  of
\begin{equation}\label{TT}
 T_{1}=T_{1}(\lambda;\lambda_1, \lambda_2)=\left( \begin{array}{cc}
a_{2}&0 \\
0 &d_{2}\\
\end{array} \right)\lambda^{2}+\left( \begin{array}{cc}
0&b_{1}\\
c_{1} &0\\
\end{array} \right)\lambda+\left( \begin{array}{cc}
a_{0}&0 \\
0 &d_{0}\\
\end{array} \right).
\end{equation}
Here $a_{2}, d_{2}, b_{1}, c_{1}$ are undetermined functions of ($x$, $t$), which will be expressed by the
eigenfunction associated with $\lambda_1$, $\lambda_2$ in the Kundu-DNLS spectral problem.
First of all, we introduce  $n$ eigenfunctions $\psi_j$ as
\begin{eqnarray}
&&\psi_{j}=\left(
\begin{array}{c}\label{jie2}
 \phi_{j}   \\
 \varphi_{j}  \\
\end{array} \right),\ \ j=1,2,....n,\phi_{j}=\phi_{j}(x,t,\lambda_{j}), \
\varphi_{j}=\varphi_{j}(x,t,\lambda_{j}). \label{jie1}
\end{eqnarray}

The  elements of one-fold Darboux transformation  are  parameterized by the
eigenfunction $\psi_1,\psi_2$ associated with
$\lambda_1, \lambda_2$ as
 \begin{eqnarray}
d_{2}&=&\dfrac{1}{a_{2}}, \ \ a_{2}=\frac{\varphi_1\phi_2\lambda_1-\phi_1\varphi_2\lambda_2}{\phi_1\varphi_2\lambda_1-\varphi_1\phi_2\lambda_2},
\ \ b_{1}=\frac{\phi_1\phi_2(\lambda_1^{2}-\lambda_2^{2})}{\varphi_1\phi_2\lambda_2-\phi_1\varphi_2\lambda_1}, \\\nonumber\ \ c_{1}&=&\frac{\varphi_1\varphi_2(\lambda_1^{2}-\lambda_2^{2})}{\phi_1\varphi_2\lambda_2-\varphi_1\phi_2\lambda_1},
\ \ a_{0}=d_{0}=\lambda_{1}\lambda_{2}.
\end{eqnarray}

Note that $(a_{2}d_{2})_{x}=0$ is derived from the eq.(\ref{xx1}), Then we
take $a_{2}=\dfrac{1}{d_{2}}$ in the followings. By transformation eq.(\ref{xx1}), new solutions are given by
 \begin{eqnarray} \label{TT1}
Q^{[1]}=\dfrac{d_{2}}{a_{2}}Q-\dfrac{c_{1}e^{-i\theta}}{a_{2}\sqrt{\alpha}}, \ \ R^{[1]}=\dfrac{a_{2}}{d_{2}}R+\dfrac{b_{1}e^{i\theta}}{d_{2}\sqrt{\alpha}}.
\end{eqnarray}

\subsection{N-fold Darboux transformation}
After considering the One-fold Darboux transformation, let us briefly discuss the N-fold Darboux transformation\cite{huang2}. First of all, we are in a position to consider the reduction of the Darboux transformation of the Kundu-DNLS equation so that $Q^{[n]}=-(R^{[n]})^*$. Under this reduction condition,
the eigenfunction $\psi_k=\left( \begin{array}{c}
\phi_k\\
\varphi_k
\end{array} \right)$ associated with eigenvalue $\lambda_k$ has following properties \cite{kenji1},

$${\phi_{2k+1}}^{\ast}=\varphi_{2k}, \ {\varphi_{2k+1}}^{\ast}=\phi_{2k},
 {\lambda_{2k+1}}^{\ast}=\lambda_{2k},  k=1,2,....n. $$
Then the Darboux transformation of the DNLS equation is given.
Now, the key task is to establish the determinant representation of the n-fold Darboux transformation for Kundu-DNLS  system in this subsection. To this purpose, set
$$
\begin{array}{cc}
\textbf{D}=&\left\{\left.\left(\begin{array}{cc}
a& 0\\
0&d
  \end{array} \right)\right| a,d \text{ are complex functions of}\ x\ \text{and}\ t  \right\},\\
%%%%%%%%%%%%%%%%%%%%%%%%%%%%%%%%%%%%%%%%%%
\textbf{A}=&\left\{\left.\left(\begin{array}{cc}
0& b\\
c&0
  \end{array} \right)\right| b,c \text{ are complex functions of}\ x\ \text{and}\ t  \right\}.
\end{array}
$$
 According to the form of $T_1$ in eq.(\ref{TT}), the n-fold Darboux transformation can be represented as
\begin{eqnarray}\label{tnss}T_{n}&=&T_{n}(\lambda;\lambda_1,\lambda_2, \cdots,\lambda_{2n})
=\sum_{i=0}^{2n}P_{i}\lambda^{i},
\end{eqnarray}
with \begin{eqnarray*}
\label{tnsss}
P_{2n}=\left( \mbox{\hspace{-0.2cm}}
\begin{array}{cc}
a_{2n}\mbox{\hspace{-0.3cm}}&0 \\
0 \mbox{\hspace{-0.3cm}}&d_{2n}\\
\end{array}  \mbox{\hspace{-0.2cm}}\right)\in \textbf{D},\ P_{2n-1}=\left( \mbox{\hspace{-0.2cm}}
\begin{array}{cc}
0 \mbox{\hspace{-0.3cm}}&b_{2n-1} \\
c_{2n-1} \mbox{\hspace{-0.3cm}}&0\\
\end{array}  \mbox{\hspace{-0.2cm}}\right)\in \textbf{A}.
\end{eqnarray*}
Here $P_{0}$ is a constant matrix, $P_i(1\leq i\leq 2n)$ is function of $x$ and $t$.

\noindent

The n-fold Darboux transformation of the Kundu-DNLS system can be expressed by
\begin{eqnarray}
\label{fss1}T_{n}&=&T_{n}(\lambda;\lambda_1,\lambda_2,\cdots,\lambda_{2n})=\left(
\begin{array}{cc}
\dfrac{{(T_{n})_{11}}}{\Delta_{n}}& \dfrac{{(T_{n})_{12}}}{\Delta_{n}}\\ \\
\dfrac{{(T_{n})_{21}}}{\widetilde{\Delta_{n}}}& \dfrac{{(T_{n})_{22}}}{\widetilde{\Delta_{n}}}\\
\end{array} \right),
\end{eqnarray}
with
\begin{equation}\label{fsst1}
\Delta_{n}=\begin{vmatrix}
\lambda_{1}^{2n}\phi_{1}&\lambda_{1}^{2n-1}\varphi_{1}&\lambda_{1}^{2n-2}\phi_{1}&\lambda_{1}^{2n-3}\varphi_{1}&\ldots&\lambda_{1}^{2}\phi_{1}&\lambda_{1}\varphi_{1}\\
\lambda_{2}^{2n}\phi_{2}&\lambda_{2}^{2n-1}\varphi_{2}&\lambda_{2}^{2n-2}\phi_{2}&\lambda_{2}^{2n-3}\varphi_{2}&\ldots&\lambda_{2}^{2}\phi_{2}&\lambda_{2}\varphi_{2}\\
\vdots&\vdots&\vdots&\vdots&\vdots&\vdots&\vdots\\
\lambda_{2n}^{2n}\phi_{2n}&\lambda_{2n}^{2n-1}\varphi_{2n}&\lambda_{2n}^{2n-2}\phi_{2n}&\lambda_{2n}^{2n-3}\varphi_{2n}&\ldots&\lambda_{2n}^{2}\phi_{2n}&\lambda_{2n}\varphi_{2n}\nonumber\\
\end{vmatrix},
\end{equation}
\begin{equation}\label{fsst2}
{(T_{n})_{11}}=\begin{vmatrix}
\lambda^{2n}&0&\lambda^{2n-2}&0&\ldots&\lambda^{2}&0&\lambda_{1}\lambda_{2}\ldots\lambda_{2n}\\
\lambda_{1}^{2n}\phi_{1}&\lambda_{1}^{2n-1}\varphi_{1}&\lambda_{1}^{2n-2}\phi_{1}&\lambda_{1}^{2n-3}\varphi_{1}&\ldots&\lambda_{1}^{2}\phi_{1}&\lambda_{1}\varphi_{1}&\lambda_{1}\lambda_{2}\ldots\lambda_{2n}\phi_{1}\\
\lambda_{2}^{2n}\phi_{2}&\lambda_{2}^{2n-1}\varphi_{2}&\lambda_{2}^{2n-2}\phi_{2}&\lambda_{2}^{2n-3}\varphi_{2}&\ldots&\lambda_{2}^{2}\phi_{2}&\lambda_{2}\varphi_{2}&\lambda_{1}\lambda_{2}\ldots\lambda_{2n}\phi_{2}\\
\vdots&\vdots&\vdots&\vdots&\vdots&\vdots&\vdots&\vdots\\
\lambda_{2n}^{2n}\phi_{2n}&\lambda_{2n}^{2n-1}\varphi_{2n}&\lambda_{2n}^{2n-2}\phi_{2n}&\lambda_{2n}^{2n-3}\varphi_{2n}&\ldots&\lambda_{2n}^{2}\phi_{2n}&\lambda_{2n}\varphi_{2n}&\lambda_{1}\lambda_{2}\ldots\lambda_{2n}\phi_{1}\nonumber\\
\end{vmatrix},
\end{equation}
\begin{equation}\label{fsst3}
{(T_{n})_{12}}=\begin{vmatrix}
0&\lambda^{2n-1}&0&\lambda^{2n-3}&\ldots&0&\lambda&0\\
\lambda_{1}^{2n}\phi_{1}&\lambda_{1}^{2n-1}\varphi_{1}&\lambda_{1}^{2n-2}\phi_{1}&\lambda_{1}^{2n-3}\varphi_{1}&\ldots&\lambda_{1}^{2}\phi_{1}&\lambda_{1}\varphi_{1}&\lambda_{1}\lambda_{2}\ldots\lambda_{2n}\phi_{1}\\
\lambda_{2}^{2n}\phi_{2}&\lambda_{2}^{2n-1}\varphi_{2}&\lambda_{2}^{2n-2}\phi_{2}&\lambda_{2}^{2n-3}\varphi_{2}&\ldots&\lambda_{2}^{2}\phi_{2}&\lambda_{2}\varphi_{2}&\lambda_{1}\lambda_{2}\ldots\lambda_{2n}\phi_{2}\\
\vdots&\vdots&\vdots&\vdots&\vdots&\vdots&\vdots&\vdots\\
\lambda_{2n}^{2n}\phi_{2n}&\lambda_{2n}^{2n-1}\varphi_{2n}&\lambda_{2n}^{2n-2}\phi_{2n}&\lambda_{2n}^{2n-3}\varphi_{2n}&\ldots&\lambda_{2n}^{2}\phi_{2n}&\lambda_{2n}\varphi_{2n}&\lambda_{1}\lambda_{2}\ldots\lambda_{2n}\phi_{1}\nonumber\\
\end{vmatrix},
\end{equation}
\begin{equation}\label{fsst4}
\widetilde{\Delta_{n}}=\begin{vmatrix}
\lambda_{1}^{2n}\varphi_{1}&\lambda_{1}^{2n-1}\phi_{1}&\lambda_{1}^{2n-2}\varphi_{1}&\lambda_{1}^{2n-3}\phi_{1}&\ldots&\lambda_{1}^{2}\varphi_{1}&\lambda_{1}\phi_{1}\\
\lambda_{2}^{2n}\varphi_{2}&\lambda_{2}^{2n-1}\phi_{2}&\lambda_{2}^{2n-2}\varphi_{2}&\lambda_{2}^{2n-3}\phi_{2}&\ldots&\lambda_{2}^{2}\varphi_{2}&\lambda_{2}\phi_{2}\\
\vdots&\vdots&\vdots&\vdots&\vdots&\vdots&\vdots\\
\lambda_{2n}^{2n}\varphi_{2n}&\lambda_{2n}^{2n-1}\phi_{2n}&\lambda_{2n}^{2n-2}\varphi_{2n}&\lambda_{2n}^{2n-3}\phi_{2n}&\ldots&\lambda_{2n}^{2}\varphi_{2n}&\lambda_{2n}\phi_{2n}\nonumber\\
\end{vmatrix},
\end{equation}
\begin{equation}\label{fsst5}
{(T_{n})_{21}}=\begin{vmatrix}
0&\lambda^{2n-1}&0&\lambda^{2n-3}&\ldots&0&\lambda&0\\
\lambda_{1}^{2n}\varphi_{1}&\lambda_{1}^{2n-1}\phi_{1}&\lambda_{1}^{2n-2}\varphi_{1}&\lambda_{1}^{2n-3}\phi_{1}&\ldots&\lambda_{1}^{2}\varphi_{1}&\lambda_{1}\phi_{1}&\lambda_{1}\lambda_{2}\ldots\lambda_{2n}\varphi_{1}\\
\lambda_{2}^{2n}\varphi_{2}&\lambda_{2}^{2n-1}\phi_{2}&\lambda_{2}^{2n-2}\varphi_{2}&\lambda_{2}^{2n-3}\phi_{2}&\ldots&\lambda_{2}^{2}\varphi_{2}&\lambda_{2}\phi_{2}&\lambda_{1}\lambda_{2}\ldots\lambda_{2n}\varphi_{2}\\
\vdots&\vdots&\vdots&\vdots&\vdots&\vdots&\vdots&\vdots\\
\lambda_{2n}^{2n}\varphi_{2n}&\lambda_{2n}^{2n-1}\phi_{2n}&\lambda_{2n}^{2n-2}\varphi_{2n}&\lambda_{2n}^{2n-3}\phi_{2n}&\ldots&\lambda_{2n}^{2}\varphi_{2n}&\lambda_{2n}\phi_{2n}&\lambda_{1}\lambda_{2}\ldots\lambda_{2n}\varphi_{1}\nonumber\\
\end{vmatrix},
\end{equation}
\begin{equation}\label{fsst6}
{(T_{n})_{22}}=\begin{vmatrix}
\lambda^{2n}&0&\lambda^{2n-2}&0&\ldots&\lambda^{2}&0&\lambda_{1}\lambda_{2}\ldots\lambda_{2n}\\
\lambda_{1}^{2n}\varphi_{1}&\lambda_{1}^{2n-2}\phi_{1}&\lambda_{1}^{2n-2}\varphi_{1}&\lambda_{1}^{2n-3}\phi_{1}&\ldots&\lambda_{1}^{2}\varphi_{1}&\lambda_{1}\phi_{1}&\lambda_{1}\lambda_{2}\ldots\lambda_{2n}\varphi_{1}\\
\lambda_{2}^{2n}\varphi_{2}&\lambda_{2}^{2n-1}\phi_{2}&\lambda_{2}^{2n-2}\varphi_{2}&\lambda_{2}^{2n-3}\phi_{2}&\ldots&\lambda_{2}^{2}\varphi_{2}&\lambda_{2}\phi_{2}&\lambda_{1}\lambda_{2}\ldots\lambda_{2n}\varphi_{2}\\
\vdots&\vdots&\vdots&\vdots&\vdots&\vdots&\vdots&\vdots\\
\lambda_{2n}^{2n}\varphi_{2n}&\lambda_{2n}^{2n-1}\phi_{2n}&\lambda_{2n}^{2n-2}\varphi_{2n}&\lambda_{2n}^{2n-3}\phi_{2n}&\ldots&\lambda_{2n}^{2}\varphi_{2n}&\lambda_{2n}\phi_{2n}&\lambda_{1}\lambda_{2}\ldots\lambda_{2n}\varphi_{1}\nonumber\\
\end{vmatrix}.
\end{equation}

Next, we consider the transformed new solutions ($Q^{[n]},R^{[n]}$)of Kundu-DNLS system corresponding to
the n-fold Kundu-DNLS transformation. Under covariant requirement of spectral problem of the Kundu-DNLS system, the transformed
form should be
\begin{eqnarray}\label{nsys11}
\partial_{x}\psi^{[n]}&=&(J\lambda^2+Q^{[n]}\lambda)\psi=U^{[n]}\psi,
\end{eqnarray}
with
\begin{equation}\label{nfj1}
     \psi=\left( \begin{array}{c}
          \phi   \\
          \varphi \\
      \end{array} \right),
    \quad J= \left( \begin{array}{cc}
       i &0 \\
       0 &-i\\
     \end{array} \right),
  \quad Q^{[n]}=\left( \begin{array}{cc}
    0 &Q^{[n]} \\
    R^{[n]} &0\\
\end{array} \right),
\end{equation}
and then  \begin{eqnarray}\label{ntt2} {T_{n}}_{x}+T_{n}~U&=&U^{[n]}~T_{n}.
\end{eqnarray}
Substituting $T_n$ given by eq.(\ref{tnss}) into eq.(\ref{ntt2})and comparing the
coefficients of $\lambda^{2n+1}$, it yields
\begin{eqnarray}\label{ntt3}
&&Q^{[n]}=\dfrac{d_{n}}{a_{n}}Q-\dfrac{c_{n-1}e^{-i\theta}}{a_{n}\sqrt{\alpha}},
\ \ R^{[n]}=\dfrac{a_{n}}{d_{n}}R+\dfrac{b_{n-1}e^{i\theta}}{d_{n}\sqrt{\alpha}}.
\end{eqnarray}
Furthermore, taking $a_{n},d_{n},b_{n-1},c_{n-1}$
which are obtained from eq.(\ref{fss1}) into (\ref{ntt3}), then new solutions ($Q^{[n]},R^{[n]}$)
are given by
\begin{eqnarray}\label{ntt4}
&&Q^{[n]}=\dfrac{\Omega_{21}^{2}}{\Omega_{11}^{2}}Q+\dfrac{e^{-i\theta}}{\sqrt{\alpha}}\dfrac{\Omega_{21}\Omega_{22}}{\Omega_{11}^{2}}, \ \ R^{[n]}=\dfrac{\Omega_{11}^{2}}{\Omega_{21}^{2}}R-\dfrac{e^{i\theta}}{\sqrt{\alpha}}\dfrac{\Omega_{11}\Omega_{12}}{\Omega_{21}^{2}}.
\end{eqnarray}

Here,
\begin{equation}\label{ntt5}
\Omega_{11}=\begin{vmatrix}
\lambda_{1}^{2n-1}\varphi_{1}&\lambda_{1}^{2n-2}\phi_{1}&\lambda_{1}^{2n-3}\varphi_{1}&\ldots&\lambda_{1}\varphi_{1}&\phi_{1}\\
\lambda_{2}^{2n-1}\varphi_{2}&\lambda_{2}^{2n-2}\phi_{2}&\lambda_{2}^{2n-3}\varphi_{2}&\ldots&\lambda_{2}\varphi_{2}&\phi_{2}\\
\vdots&\vdots&\vdots&\vdots&\vdots&\vdots\\
\lambda_{2n}^{2n-1}\varphi_{2n}&\lambda_{2n}^{2n-2}\phi_{2n}&\lambda_{2n}^{2n-3}\varphi_{2n}&\ldots&\lambda_{2n}\varphi_{2n}&\phi_{2n}\\
\end{vmatrix},
\end{equation}
\begin{equation*}
\Omega_{12}=\begin{vmatrix}
\lambda_{1}^{2n}\phi_{1}&\lambda_{1}^{2n-2}\phi_{1}&\lambda_{1}^{2n-3}\varphi_{1}&\ldots&\lambda_{1}\varphi_{1}&\phi_{1}\\
\lambda_{2}^{2n}\phi_{2}&\lambda_{2}^{2n-2}\phi_{2}&\lambda_{2}^{2n-3}\varphi_{2}&\ldots&\lambda_{2}\varphi_{2}&\phi_{2}\\
\vdots&\vdots&\vdots&\vdots&\vdots&\vdots\\
\lambda_{2n}^{2n}\phi_{2n}&\lambda_{2n}^{2n-2}\phi_{2n}&\lambda_{2n}^{2n-3}\varphi_{2n}&\ldots&\lambda_{2n}\varphi_{2n}&\phi_{2n}\\
\end{vmatrix},
\end{equation*}
\begin{equation*}
\Omega_{21}=\begin{vmatrix}
\lambda_{1}^{2n-1}\phi_{1}&\lambda_{1}^{2n-2}\varphi_{1}&\lambda_{1}^{2n-3}\phi_{1}&\ldots&\lambda_{1}\phi_{1}&\varphi_{1}\\
\lambda_{2}^{2n-1}\phi_{2}&\lambda_{2}^{2n-2}\varphi_{2}&\lambda_{2}^{2n-3}\phi_{2}&\ldots&\lambda_{2}\phi_{2}&\varphi_{2}\\
\vdots&\vdots&\vdots&\vdots&\vdots&\vdots\\
\lambda_{2n}^{2n-1}\phi_{2n}&\lambda_{2n}^{2n-2}\varphi_{2n}&\lambda_{2n}^{2n-3}\phi_{2n}&\ldots&\lambda_{2n}\phi_{2n}&\varphi_{2n}\\
\end{vmatrix},
\end{equation*}
\begin{equation*}
\Omega_{22}=\begin{vmatrix}
\lambda_{1}^{2n}\varphi_{1}&\lambda_{1}^{2n-2}\varphi_{1}&\lambda_{1}^{2n-3}\phi_{1}&\ldots&\lambda_{1}\phi_{1}&\varphi_{1}\\
\lambda_{1}^{2n}\varphi_{1}&\lambda_{2}^{2n-2}\varphi_{2}&\lambda_{2}^{2n-3}\phi_{2}&\ldots&\lambda_{2}\phi_{2}&\varphi_{2}\\
\vdots&\vdots&\vdots&\vdots&\vdots&\vdots\\
\lambda_{2n}^{2n}\varphi_{2n}&\lambda_{2n}^{2n-2}\varphi_{2n}&\lambda_{2n}^{2n-3}\phi_{2n}&\ldots&\lambda_{2n}\phi_{2n}&\varphi_{2n}\\
\end{vmatrix}.
\end{equation*}

So far, we discussed about the determinant construction of n-th Darboux transformation of Kundu-DNLS equation.
As an application of these transformations of Kundu-DNLS equation, soliton solutions and positon solutions
will be constructed in the next section.

%%%%%%%%%%%%%%%%%%%%%%%%%%%%%%%%%%%%%%%%%%%%%%%%%%%%%%%%55
\section{Soliton solutions and Positon solution of the Kundu-DNLS equation}

For $Q=0$ the equations (\ref{TT1}) is solved by
 \begin{eqnarray}\label{jie1}
 &&\psi_{j}=\left( \begin{array}{c}
 \phi_{j} \\
 \varphi_{j}\\
 \end{array} \right),\quad
  \phi_{j}=\exp(-\frac{i}{8}(2{\lambda_{j}}^{2} x+ {\lambda_{j}}^{4}t)), \quad  \varphi_{j}=\exp(\frac{i}{8}(2{\lambda_{j}}^{2} x+{\lambda_{j}}^{4}t)).\nonumber\\
\end{eqnarray}

Considering the choice in eq.(\ref{TT1}) with
 $\lambda_{1}=m_{1}+in_{1}$, $\lambda_{2} =\lambda_{1}^{\ast}$, $m_{1},n_{1}\in\mathbb{R}$ and using eigenfunctions
in eq.(\ref{jie1}), then one soliton solution of Kundu-DNLS equation is
\begin{eqnarray}\label{jie22}
 &&Q^{[1]}=\frac{(e^{iF_{1}}\lambda_{1}-e^{iF_{2}}\lambda_{2})({\lambda_{1}}+{\lambda_{2}})}{e^{-2if}\sqrt{\alpha}(\lambda_{1}-\lambda_{2})},\nonumber
 \\
  \end{eqnarray}
with $ F_{i}=-\frac{1}{4}(-2\lambda_{i}^{2}x+\lambda_{i}^{4}t+4\theta),i=1,2;f=\frac{1}{8}(\lambda_{1}-\lambda_{2})(\lambda_{1}+\lambda_{2})(t\lambda_{1}^{2}-2x+t\lambda_{2}^{2} ) $ .
The picture of one soliton solution of the Kundu-DNLS equation and its corresponding density graph are plotted in Fig.1 .
%%%%%%%%%%%%%%%%%%%%%%%%%%%%%
\begin{figure}[h!]
\centering
\raisebox{0.85in}{}\includegraphics[scale=0.37]{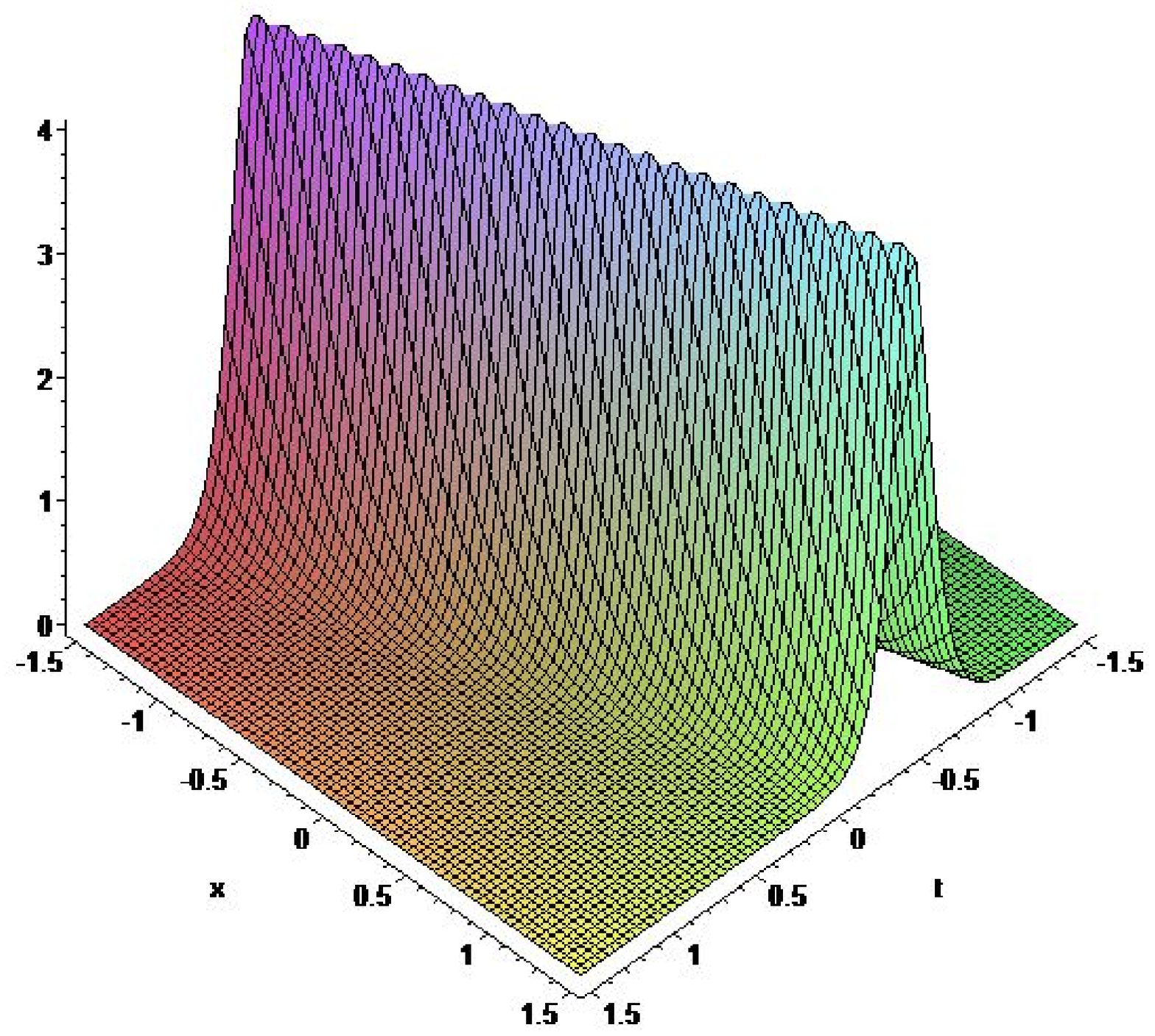}
\hskip 0.01cm
\raisebox{0.85in}{}\raisebox{-0.1cm}{\includegraphics[scale=0.27]{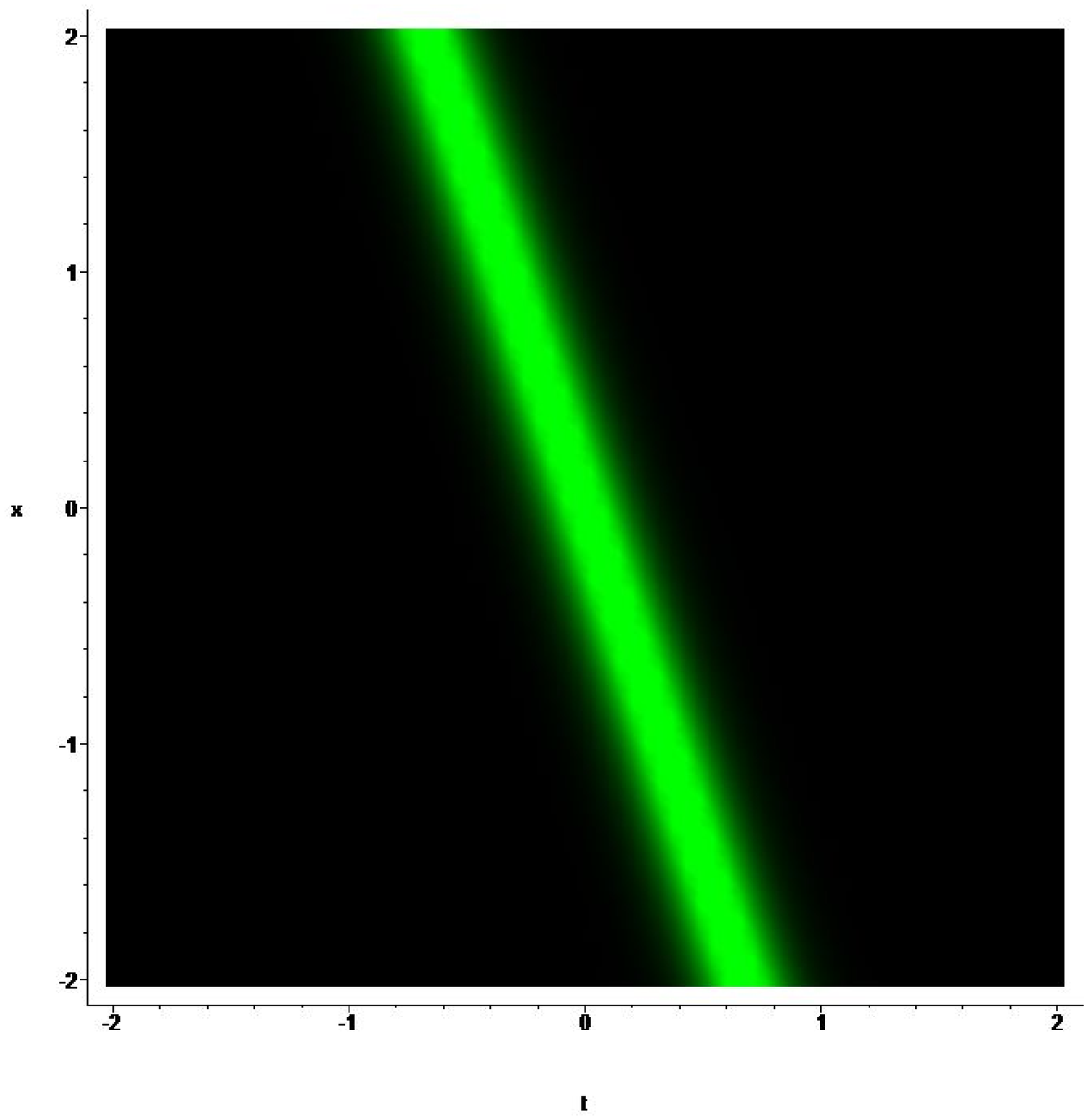}}
%\hskip 0.01cm
%\raisebox{0.85in}{($\eta$)}\raisebox{-0.1cm}{\includegraphics[scale=0.18]{V2-soliton-y}}
 \caption{One soliton solution $|Q^{[1]}|^2$ of the Kundu-DNLS equation when
 $m_{1}=1,n_{1}=2,\alpha=1,\theta=x+t.$}\label{1soliton}
\end{figure}
%%%%%%%%%%%%%%%%%%%%%%%%%%%%%%%%%%%%%%%%%%%%%%%%%%%%%%%%%%%

Now let us discuss about the construction of the two-soliton
solution of Kundu-DNLS system. For the purpose, we have to use two
spectral parameters $ \lambda_{1} = m_{1} +in_{1} $, $\lambda_{2} =\lambda_{1}^{\ast}$,and $\lambda_{3} = m_{2}+in_{2}$, $\lambda_{4} =\lambda_{3}^{\ast}$. After two Darboux transformations, the two
soliton solution is derived as follows:
\begin{eqnarray}\label{jie22}
 &&Q^{[2]}=\frac{K_1}{K_2}
  \end{eqnarray}
where
\begin{eqnarray*}
K_1&=&-4ie^{-i\theta}(2Mcos\rho_{1}+H_
1000
{1}e^{\rho_{2}}+H_{2}e^{\rho_{3}}+H_{3}e^{\rho_{4}}+H_{4}e^{\rho_{5}})\\
&&(H_{5}e^{\rho_{6}}+H_{6}e^{-\rho_{6}^{*}}+H_{7}e^{\rho_{7}}+H_{8}e^{-\rho_{7}^{*}}),\\
K_2&=&-\sqrt{\alpha}(2Mcos\rho_{1}+H_{4}e^{\rho_{2}}+H_{3}e^{\rho_{3}}+H_{2}e^{\rho_{4}}+H_{1}e^{\rho_{5}})^{2},\\
A&=&-(m_1+m_2)+i(n_1+n_2), \ \ B=(m_2-m_1)+i(n_1-n_2),\\
C&=&(m_2-m_1)+i(n_1+n_2), \ \ D=-(m_1+m_2)+i(n_1-n_2),\\
E&=&in_1-m_1, \ \ F=in_2-m_2,\\
H_{1}&=&-|A|^{2}|C|^{2}EF, \ \  H_{2}=|B|^{2}|D|^{2}EF^{*},\\
H_{3}&=&|B|^{2}|D|^{2}E^{*}F, \ \ H_{4}=-|A|^{2}|C|^{2}E^{*}F^{*},\\
H_{5}&=&-m_1n_1A^{*}BC^{*}DF, \ \ H_{6}=m_1n_1AB^{*}CDF^{*},\\
H_{7}&=&m_2n_2A^{*}B^{*}CDE, \ \ H_{8}=m_1n_1ABCD^{*}E^{*},\\
\rho_{1}&=&\frac{1}{4}(tm_1^{4}-tm_2^{4}-tn_2^{4}+tn_1^{4}+6tm_2^{2}n_2^{2}-6tm_1^{2}n_1^{2}-2xm_1^{2}+2xm_2^{2}+2xn_1^{2}-2xn_2^{2}),\\
\rho_{2}&=&tm_1^{3}n_1-tm_1n_1^{3}+tm_2^{3}n_2-tm_2n_2^{3}-xm_1n_1-xm_2n_2,\\
\rho_{3}&=&tm_1^{3}n_1-tm_1n_1^{3}-tm_2^{3}n_2+tm_2n_2^{3}-xm_1n_1+xm_2n_2,\\
\rho_{4}&=&-tm_1^{3}n_1+tm_1n_1^{3}+tm_2^{3}n_2-tm_2n_2^{3}+xm_1n_1-xm_2n_2,\\
\rho_{5}&=&-tm_1^{3}n_1+tm_1n_1^{3}-tm_2^{3}n_2+tm_2n_2^{3}+xm_1n_1+xm_2n_2,\\
\rho_{6}&=&\frac{-i}{4}(tm_1^{4}+tn_1^{4}+4itm_2^{3}n_2-4itm_2n_2^{3}-6tm_1^{2}n_1^{2}-2xm_1^{2}+2xn_1^{2}-4ixm_2n_2),\\
\rho_{7}&=&\frac{-i}{4}(tm_2^{4}+tn_2^{4}+4itm_1^{3}n_1-4itm_1n_1^{3}-6tm_2^{2}n_2^{2}+2xm_2^{2}+2xn_2^{2}-4ixm_1n_1).\\
\end{eqnarray*}
The picture of two soliton solution of the Kundu-DNLS equation and its corresponding density graph
are plotted in Fig.2 .
%%%%%%%%%%%%%%%%%%%%%%%%%%%%%
\begin{figure}[h!]
\centering
\raisebox{0.85in}{}\includegraphics[scale=0.39]{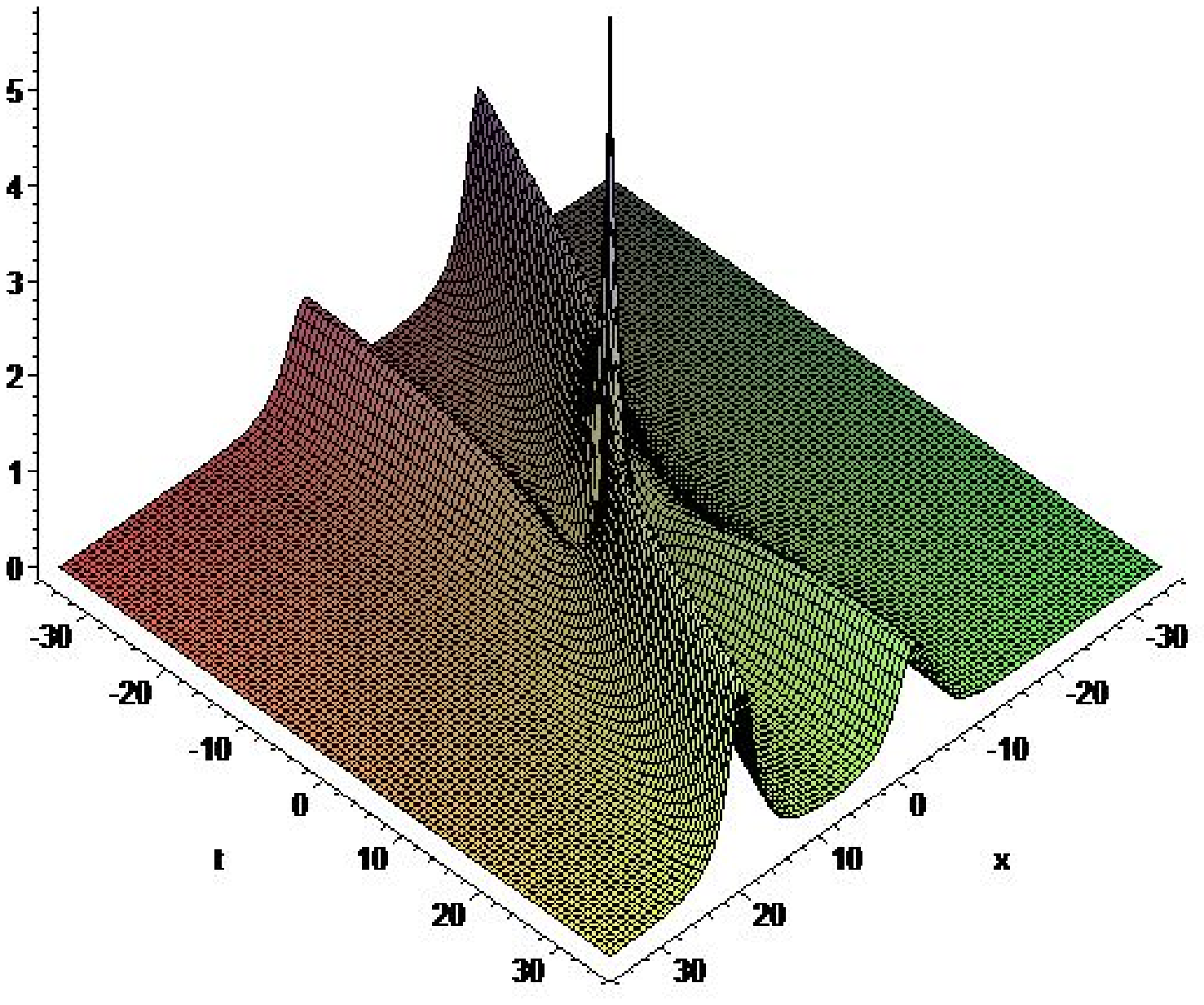}
\hskip 0.01cm
\raisebox{0.85in}{}\raisebox{-0.1cm}{\includegraphics[scale=0.25]{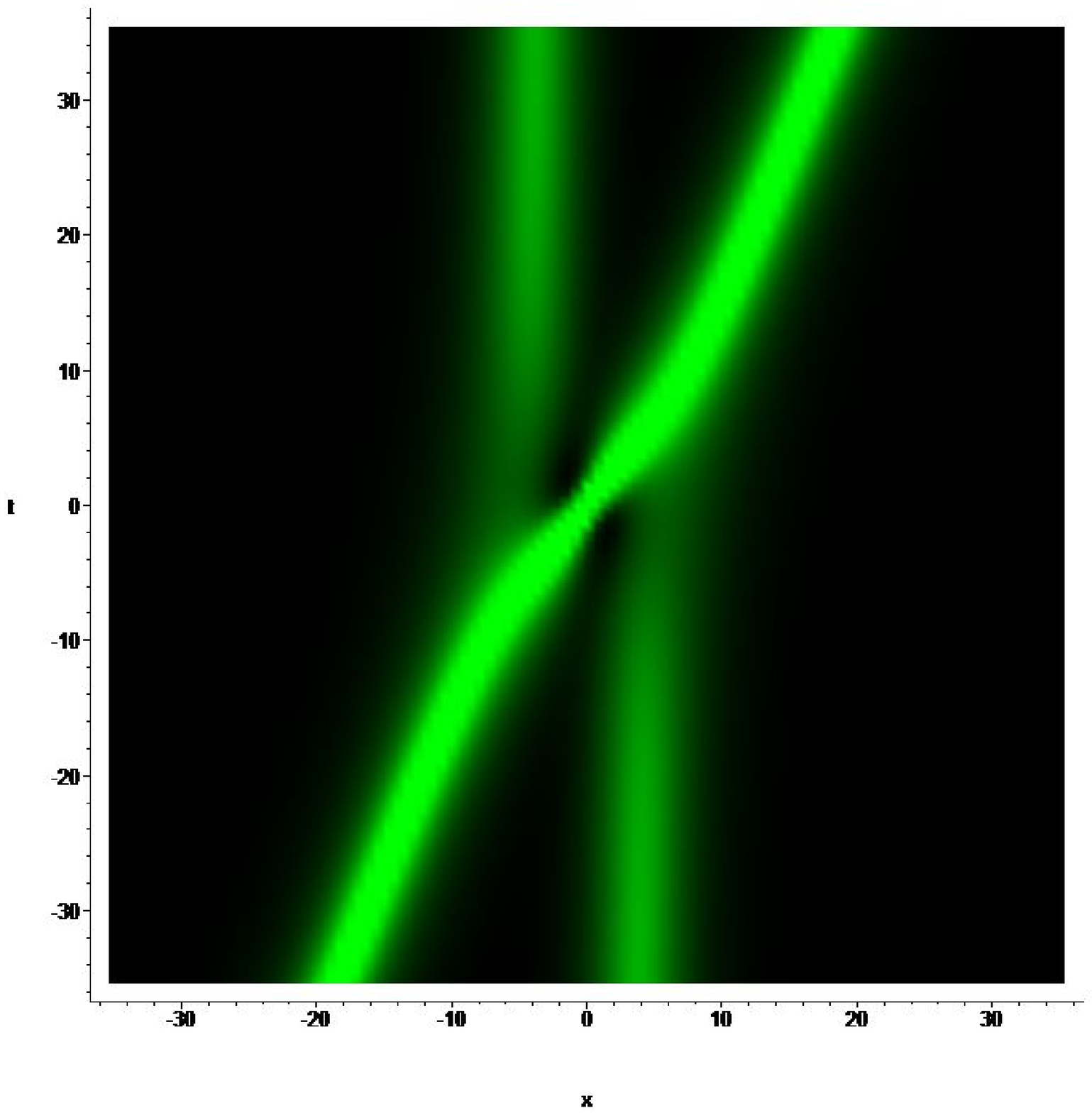}}
%\hskip 0.01cm
%\raisebox{0.85in}{($\eta$)}\raisebox{-0.1cm}{\includegraphics[scale=0.18]{V2-soliton-y}}
 \caption{Two soliton solution $|Q^{[2]}|^2$  of the Kundu-DNLS equation when
 $m_1=0.7,m_2=0.5,n_1=0.3,n_2=0.5,\alpha=1.$}\label{2soliton}
\end{figure}
%%%%%%%%%%%%%%%%%%%%%%%%%%%%%%%%%%%%%%%%%%%%%%%%%%%%%%%%%%%

\begin{figure}[h!]
\centering
\raisebox{0.85in}{}\includegraphics[scale=0.39]{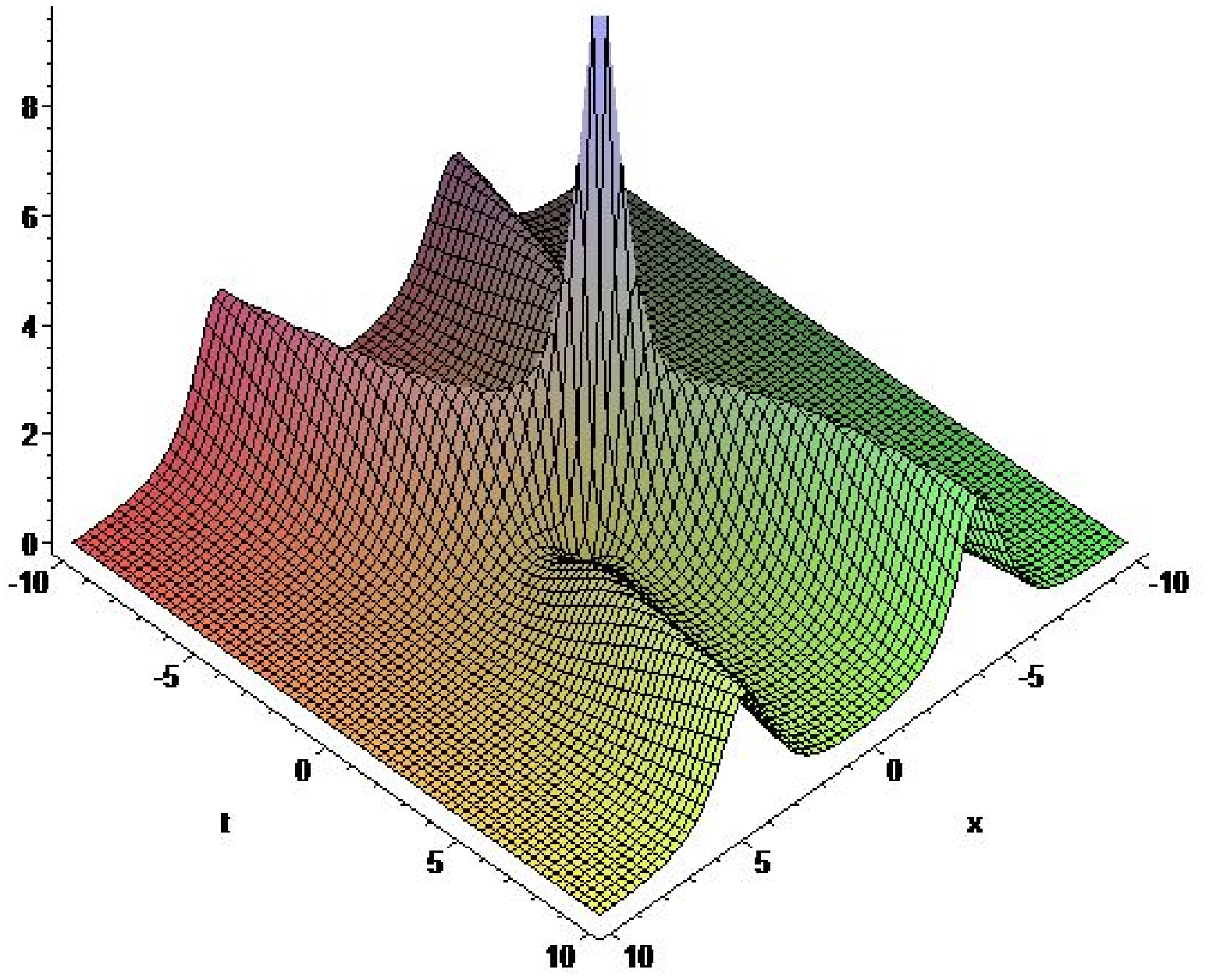}
\hskip 0.01cm
\raisebox{0.85in}{}\raisebox{-0.1cm}{\includegraphics[scale=0.25]{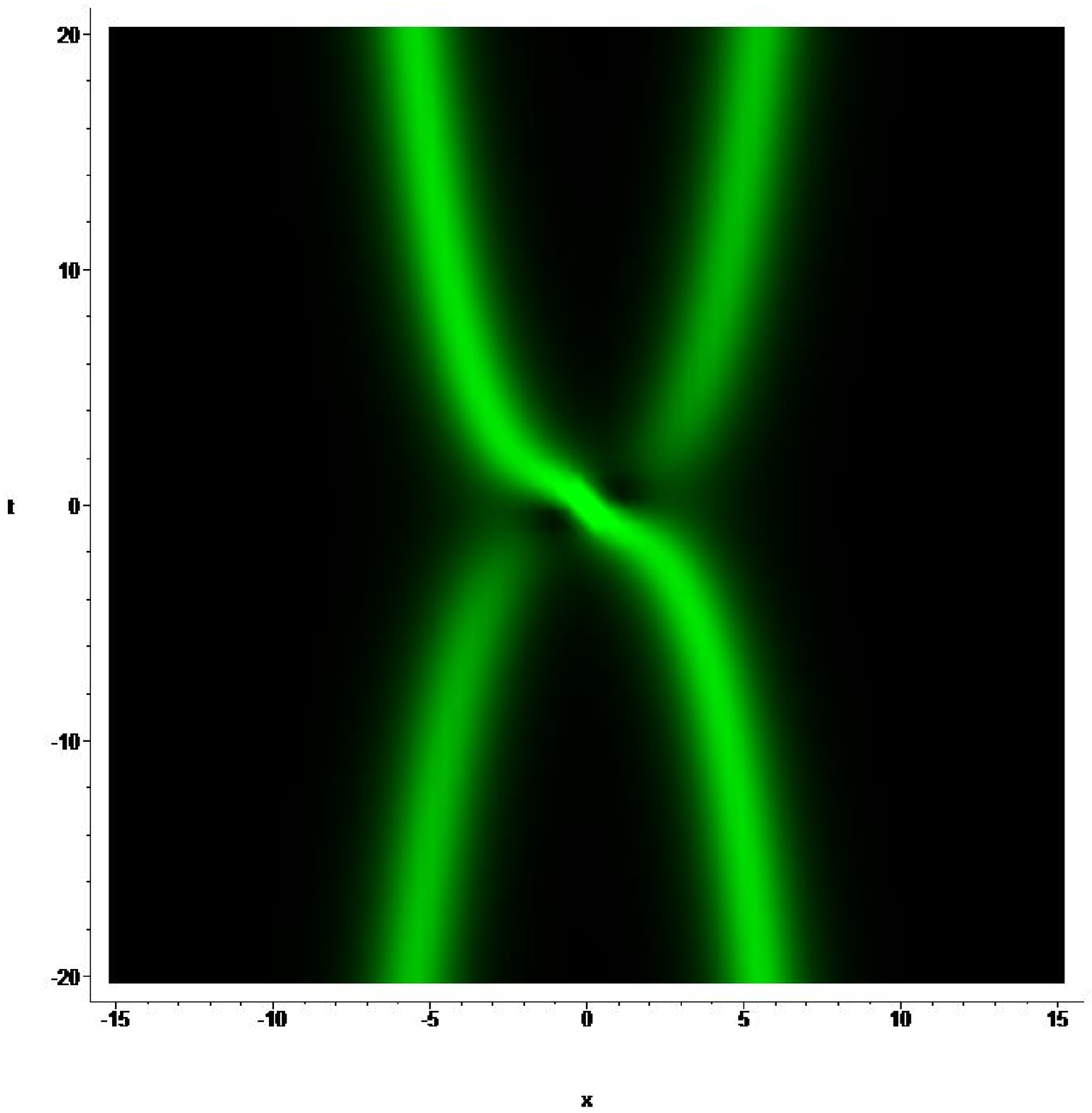}}
%\hskip 0.01cm
%\raisebox{0.85in}{($\eta$)}\raisebox{-0.1cm}{\includegraphics[scale=0.18]{V2-soliton-y}}
 \caption{Positon  solution $|Q_{p}|^2$ of the Kundu-DNLS equation with
 $\alpha_1=0.8,\beta_1=0.8.$}\label{positon1}
\end{figure}

In the following, we consider the construction of positon solution of Kundu-DNLS equation in detail. From the two soliton solution, we make use of four spectral parameters $
\lambda_{1} = \alpha_{1} +i\beta_{1} $, $ \lambda_{2} = \alpha_{1}
-i\beta_{1} $and $\lambda_{3} = \alpha_{1}+i\beta_{1}+\varepsilon$,
$\lambda_{4} = \alpha_{1} -i\beta_{1}+\varepsilon$. By letting the infinitesimal complex number $\varepsilon$ and
doing the Taylor expansion of wave function to $\lambda_{1}$, $\lambda_{2}$, positon solution as the method of ref.\cite{HJS}.
Following four linear functions are used to construct the second  positon solution,
 \begin{eqnarray}\label{jie1}
 &&\psi_{k}=\left( \begin{array}{c}
 \phi_{k} \\
 \varphi_{k}\\
 \end{array} \right),\quad
  \phi_{k}=\exp(-\frac{i}{8}(2{\lambda_{k}}^{2} x+ {\lambda_{k}}^{4}t)), \quad  \varphi_{k}=\exp(\frac{i}{8}(2{\lambda_{k}}^{2} x+{\lambda_{k}}^{4}t)),\nonumber\\
\end{eqnarray}
where $k=1,2,3,4 $. The positon solution of the Kundu-DNLS equation is given as follows:
\begin{eqnarray}\label{jie22}
 &&Q_{p}=\frac{-8\alpha_{1}\beta_{1}G_{1}G_{2}(G_{3}+G_{4})}{(G_{3}-G_{4})^{2}},\nonumber\\
\end{eqnarray}
where
\begin{eqnarray*}
G_{1}&=&i\alpha_{1}^{3}\cosh(\alpha_{1}\beta_{1}g_{1})+2\alpha_{1}^{3}\beta_{1}t\cosh(\alpha_{1}\beta_{1}g_{1})-\alpha_{1}^{3}\beta_{1}^{2}x\cosh(\alpha_{1}\beta_{1}g_{1})\\
&&-\alpha_{1}\beta_{1}^{4}x\cosh(\alpha_{1}\beta_{1}g_{1})-\alpha_{1}\beta_{1}^{6}t\cosh(\alpha_{1}\beta_{1}g_{1})+3\alpha_{1}^{5}\beta_{1}^{2}t\cosh(\alpha_{1}\beta_{1}g_{1})\\
&&-i\alpha_{1}^{6}\beta_{1}t\sinh(\alpha_{1}\beta_{1}g_{1})+2i\alpha_{1}^{4}\beta_{1}^{3}t\sinh(\alpha_{1}\beta_{1}g_{1})+i\alpha_{1}^{4}\beta_{1}x\sinh(\alpha_{1}\beta_{1}g_{1})\\
&&+i\alpha_{1}^{2}\beta_{1}^{3}x\sinh(\alpha_{1}\beta_{1}g_{1})+3i\alpha_{1}^{2}\beta_{1}^{2}t\sinh(\alpha_{1}\beta_{1}g_{1})-\beta_{1}^{3}\sinh(\alpha_{1}\beta_{1}g_{1}),\\
G_{2}&=&\cos H_{2}+i \sin H_{2},\\
G_{3}&=&2i\alpha_{1}^{3}\beta_{1}\sinh(2\alpha_{1}\beta_{1}g_{1})+4i\alpha_{1}^{2}\beta_{1}^{6}t-4i\alpha_{1}^{4}\beta_{1}^{2}x-24i\alpha_{1}^{4}\beta_{1}^{4}t+\\
&&2i\alpha_{1}\beta_{1}^{3}\sinh(2\alpha_{1}\beta_{1
2000
}g_{1})+4i\alpha_{1}^{6}\beta_{1}^{2}t+4i\alpha_{1}^{2}\beta_{1}^{2}t+4i\alpha_{1}^{2}\beta_{1}^{4}x,\\
G_{4}&=&\alpha_{1}^{4}+\beta_{1}^{4}-4\alpha_{1}^{8}\beta_{1}^{2}xt-4\alpha_{1}^{4}\beta_{1}^{6}xt-4\alpha_{1}^{6}\beta_{1}^{4}xt+4\alpha_{1}^{2}\beta_{1}^{8}xt+4\alpha_{1}^{4}\beta_{1}^{4}x^{2}\\
&&+8\alpha_{1}^{4}\beta_{1}^{8}t^{2}+2\alpha_{1}^{6}\beta_{1}^{2}x^{2}+2\alpha_{1}^{10}\beta_{1}^{2}t^{2}+8\alpha_{1}^{8}\beta_{1}^{4}t^{2}+12\alpha_{1}^{6}\beta_{1}^{6}t^{2}+2\alpha_{1}^{2}\beta_{1}^{6}x^{2}\\
&&+2\alpha_{1}^{2}\beta_{1}^{10}t^{2}-\beta_{1}^{4}\cosh(2\alpha_{1}\beta_{1}g_{1})+\alpha_{1}^{4}\cosh(2\alpha_{1}\beta_{1}g_{1}),\\
g_{1}&=&-x-t\alpha_{1}^{2}-t\beta_{1}^{2},\\
g_{2}&=&x+t+\frac{1}{4}t\beta_{1}^{4}-\frac{3}{2}t\alpha_{1}^{2}\beta_{1}^{2}+\frac{1}{2}x\beta_{1}^{2}+\frac{1}{4}t\alpha_{1}^{4}-\frac{1}{2}x\alpha_{1}^{2}.\\
\end{eqnarray*}
The picture of positon solution of the Kundu-DNLS equation and its corresponding density graph
are plotted in Fig.\ref{positon1}. When $t\rightarrow \infty$, one can find that the difference between the two  soliton solution in Fig.\ref{2soliton} and the positon solution in Fig.\ref{positon1} as following.
Positon can be treated as one special case of two solitons. When two spectral parameters of  two  soliton solution get closer and closer, the separating speed between two branches of two solitons becomes slower and slower. This approximation leads two soliton solutions to one Positon solutions. That is why Positon solutions are also called degenerated solitons. Positons of Kundu-DNLS equation are in fact long-range analogues of solitons of Kundu-DNLS equation and are slowly decreasing, oscillating solutions.

%%%%%%%%%%%%%%%%%%%%%%%%%%%%%
%%%%%%%%%%%%%%%%%%%%%%%%%%%%%%%%%%%%%%%%%%%%%%%%%%%%%%%%%%%
\section{Breather solution and Rogue wave solutions of the Kundu-DNLS equation}

From last two sections, soliton and positon solutions have been
generated for the Kundu-DNLS equation through the Darboux transformation. However, these solutions were
obtained from a trivial seed-zero solution. In this
section, we derive the new kind of solution (that is breather
solution) from a periodic seed solution.

Taking the seed solution as $Q=ce^{i\rho}, \theta=x+t, \rho=ax+bt$, which admits the constraint in the
form
\begin{eqnarray}
b = -\alpha c^2 a-2-a^2-2a-\alpha c^2.
\end{eqnarray}
Then the eigenfunction conrresponded to eigenvalue $\lambda$ is obtained.
Using the method of separation of variables and the superposition
principle, the eigenfunction $\psi_k$ associated with $\lambda_k$ is given by
\begin{eqnarray}\label{eigenfunfornonzeroseed}
\left(\mbox{\hspace{-0.2cm}} \begin{array}{c}
 \phi_{k}(x,t,\lambda_{k})\\
 \varphi_{k}(x,t,\lambda_{k})\\
\end{array}\mbox{\hspace{-0.2cm}}\right)\mbox{\hspace{-0.2cm}}=\mbox{\hspace{-0.2cm}}\left(\mbox{\hspace{-0.2cm}}\begin{array}{c}
f_1(x,t,\lambda_{k})[1,k]+f_2(x,t,\lambda_{k})[1,k]+f_1^{\ast}(x,t,{\lambda_{k}^{\ast})}[2,k]+f_2^{\ast}(x,t,{\lambda_{k}^{\ast})}[2,k]\\
f_1(x,t,\lambda_{k})[2,k]+f_2(x,t,\lambda_{k})[2,k]+f_1^{\ast}(x,t,{\lambda_{k}^{\ast})}[1,k]+f_2^{\ast}(x,t,{\lambda_{k}^{\ast})}[1,k]\\
\end{array}\mbox{\hspace{-0.2cm}}\right).
\end{eqnarray}
Here
\begin{eqnarray*}
\left(\mbox{\hspace{-0.2cm}}\begin{array}{c}
 f_1(x,t,\lambda_{k})[1,k]\\
 f_1(x,t,\lambda_{k})[2,k]\\
\end{array}\mbox{\hspace{-0.2cm}}\right)\mbox{\hspace{-0.2cm}}=\mbox{\hspace{-0.2cm}}\left(\mbox{\hspace{-0.2cm}}\begin{array}{c}
{\frac { 2-{\lambda}^{2}+2\,a-s }{2\lambda\,c}} {{\rm e}^{\frac{1}{8}i\left( -4\,ax-4\,x-2\,xs+t{\lambda}^{2}s+8\,ta+4\,t{a}^{2}+2\,tas+4\,
t+2\,ts+4\,t{c}^{2}+4\,t{c}^{2}a+2\,t{c}^{2}s \right) }}\\
{{\rm e}^{\frac{1}{8}i \left( 4\,ax+4\,x-2\,xs+t{\lambda}^{2}s-8\,ta-4\,t{a}
^{2}+2\,tas-4\,t+2\,ts-4\,t{c}^{2}-4\,t{c}^{2}a+2\,t{c}^{2}s \right) }} \\
\end{array}\mbox{\hspace{-0.2cm}}\right),\\
\end{eqnarray*}
\begin{eqnarray*}
\left(\mbox{\hspace{-0.2cm}}\begin{array}{c}
 f_2(x,t,\lambda_{k})[1,k]\\
 f_2(x,t,\lambda_{k})[2,k]\\
\end{array}\mbox{\hspace{-0.2cm}} \right)\mbox{\hspace{-0.2cm}}=\mbox{\hspace{-0.3cm}}
\left(\mbox{\hspace{-0.3cm}}\begin{array}{c}
{\frac {  2-{\lambda}^{2}+2\,a+s }{2\lambda\,c}} {{\rm e}^{\frac{-1}{8}i\left( 4\,ax+4\,x-2\,xs+t{\lambda}^{2}s-8\,ta-4\,t{a}^{2}+2\,tas-4\,t
+2\,ts-4\,t{c}^{2}-4\,t{c}^{2}a+2\,t{c}^{2}s \right) }}\\
{{\rm e}^{\frac{-1}{8}i \left( -4\,ax-4\,x-2\,xs+t{\lambda}^{2}s+8\,ta+4\,t{
a}^{2}+2\,tas+4\,t+2\,ts+4\,t{c}^{2}+4\,t{c}^{2}a+2\,t{c}^{2}s
 \right) }}\\
\end{array}\mbox{\hspace{-0.3cm}}\right)\mbox{\hspace{-0.1cm}},\\
\end{eqnarray*}

\begin{eqnarray*}
f_1(x,t,\lambda_{k})=
\left( \begin{array}{c}
 f_1(x,t,\lambda_{k})[1,k]\\
 f_1(x,t,\lambda_{k})[2,k]\\
\end{array} \right),~~~~~
f_2(x,t,\lambda_{k})=
\left( \begin{array}{c}
 f_2(x,t,\lambda_{k})[1,k]\\
 f_2(x,t,\lambda_{k})[2,k]\\
\end{array} \right),
\end{eqnarray*}
\begin{eqnarray}
&s=\sqrt{4a^{2}-4a{\lambda}^{2}+8a+{\lambda}^{4}-4{\lambda}^{2}+4-4{\lambda}^{2}c^{2}}. \nonumber
\end{eqnarray}

Note that
$f_1(x,t,\lambda_{k})$ and $f_2(x,t,\lambda_{k})$ are two different solutions of
the spectral problem eq.(\ref{ABC1}),
but we  can only get the  trivial solutions through DT of the Kundu-DNLS equation by setting eigenfunction $\psi_k$ to be one of them.

Now let us discuss about the construction of the breather solution of Kundu-DNLS equation. For the purpose, we have to use two
spectral parameters $ \lambda_{1} =\xi +i\eta, $ and $ \lambda_{2} =\xi -i\eta.$
To simplify the calculations, we use the second Darboux transformation discussed in the last section, then the breather solution $Q_{b}$ with $a=-2,c=1,\xi=0.5,\eta=1$ is obtained in the form
\begin{eqnarray}
&&Q_{b}=\frac{-b_{1}b_{2}}{2b_{3}^{2}},\nonumber\\
\end{eqnarray}
where
\begin{eqnarray*}
b_{1}&=&63508327ie^{-0.9682458364ix}+436491673e^{0.2420614592t}-563508327ie^{0.2420614592t}\\
&&-436491673e^{-0.2420614592t}-563508327ie^{-0.2420614592t}+5\times10^{8}ie^{-0.9682458364ix},\\
b_{2}&=&1309475019e^{-0.2420614592t-2ix-it}+10\times10^{8}ie^{-0.4\times10^{-8}i(257938541x+2.5\times10^{8}t)}\\
&&-1309475019e^{0.2420614592t-2ix-it}+563508327ie^{0.2420614592t-2ix-it}\\
&&+563508327e^{-0.2420614592t-2ix-it}+127016654ie^{-0.4\times10^{-8}i(742061459x+2.5\times10^{8}t)},\\
b_{3}&=&5\times10^{8}ie^{-0.9682458364ix}+63508327ie^{0.9682458364ix}-436491673e^{0.2420614592t}\\
&&-563508327ie^{0.2420614592t}+436491673e^{-0.2420614592t}-563508327ie^{-0.2420614592t}.
\end{eqnarray*}
The picture of breather solutions of the Kundu-DNLS equation and its corresponding density graph are plotted in Fig.4, which propagates along the line $t=0$, however, by changing the value of the parameters,
 the direction of propagating for the breather will change.
%%%%%%%%%%%%%%%%%%%%%%%%%%%%%
\begin{figure}[h!]
\centering
\raisebox{0.85in}{}\raisebox{-0.1cm}{\includegraphics[scale=0.39]{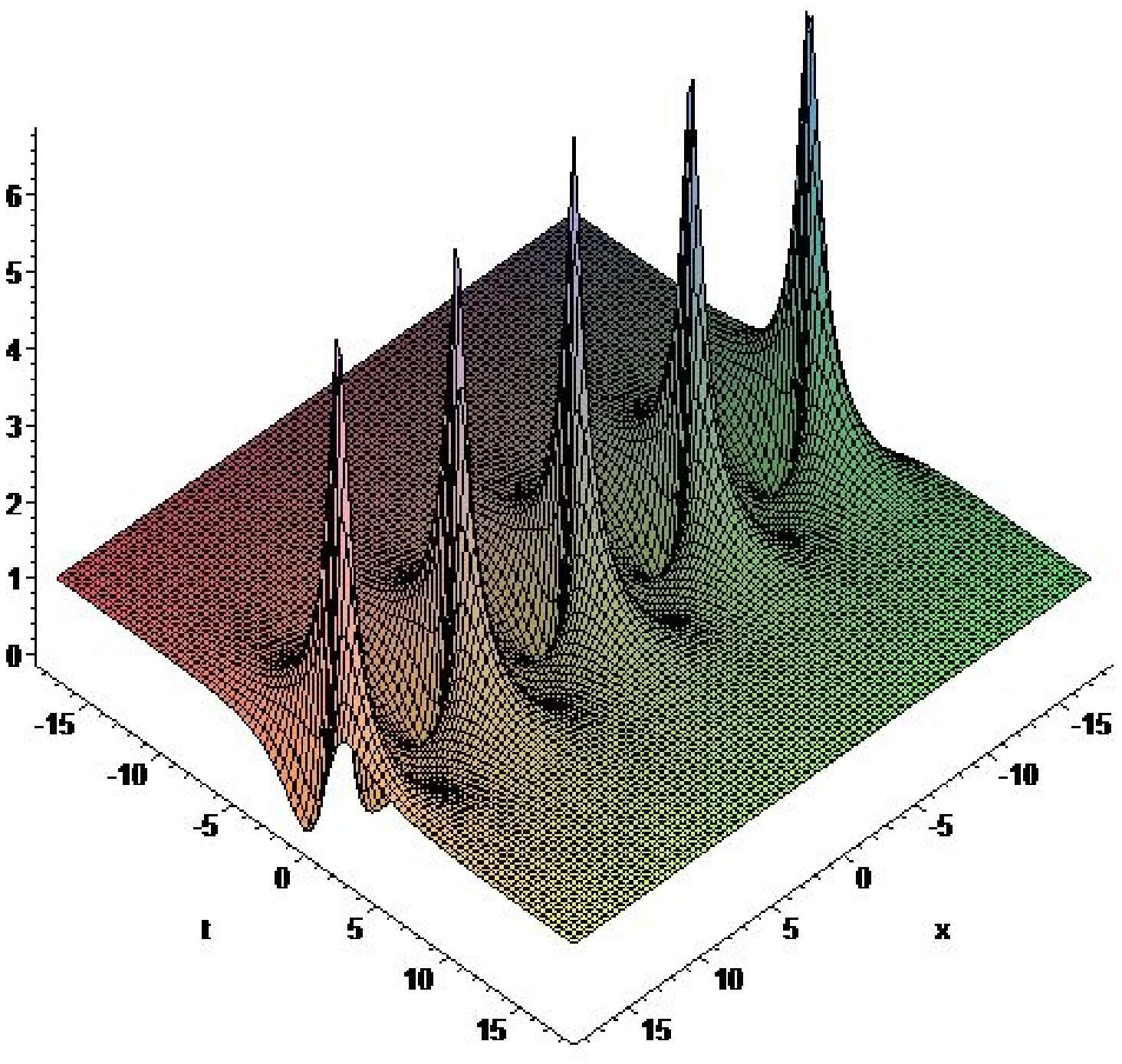}}
\hskip 0.01cm
\raisebox{0.85in}{}\raisebox{-0.1cm}{\includegraphics[scale=0.23]{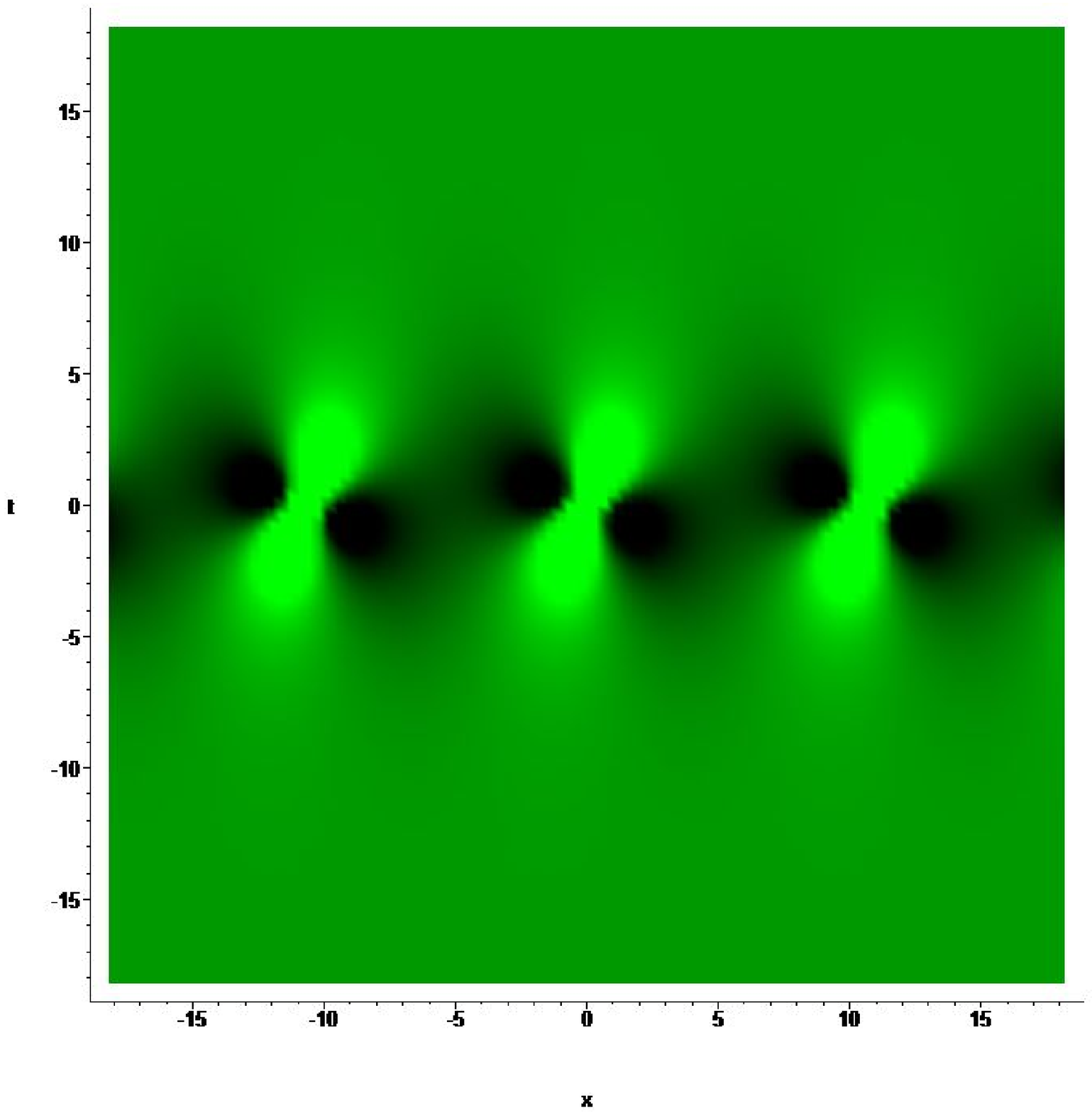}}
 \caption{Breather  solution $|Q_{b}|^2$  of the Kundu-DNLS equation with
 $a=-2,c=1,\xi=0.5,\eta=1.$}\label{breather}
\end{figure}
%%%%%%%%%%%%%%%%%%%%%%%%%%%%%%%%%%%%%%%%%%%%%%%%%%%%%%%%%%%

\begin{figure}[h!]
\centering
\raisebox{0.85in}{}\includegraphics[scale=0.39]{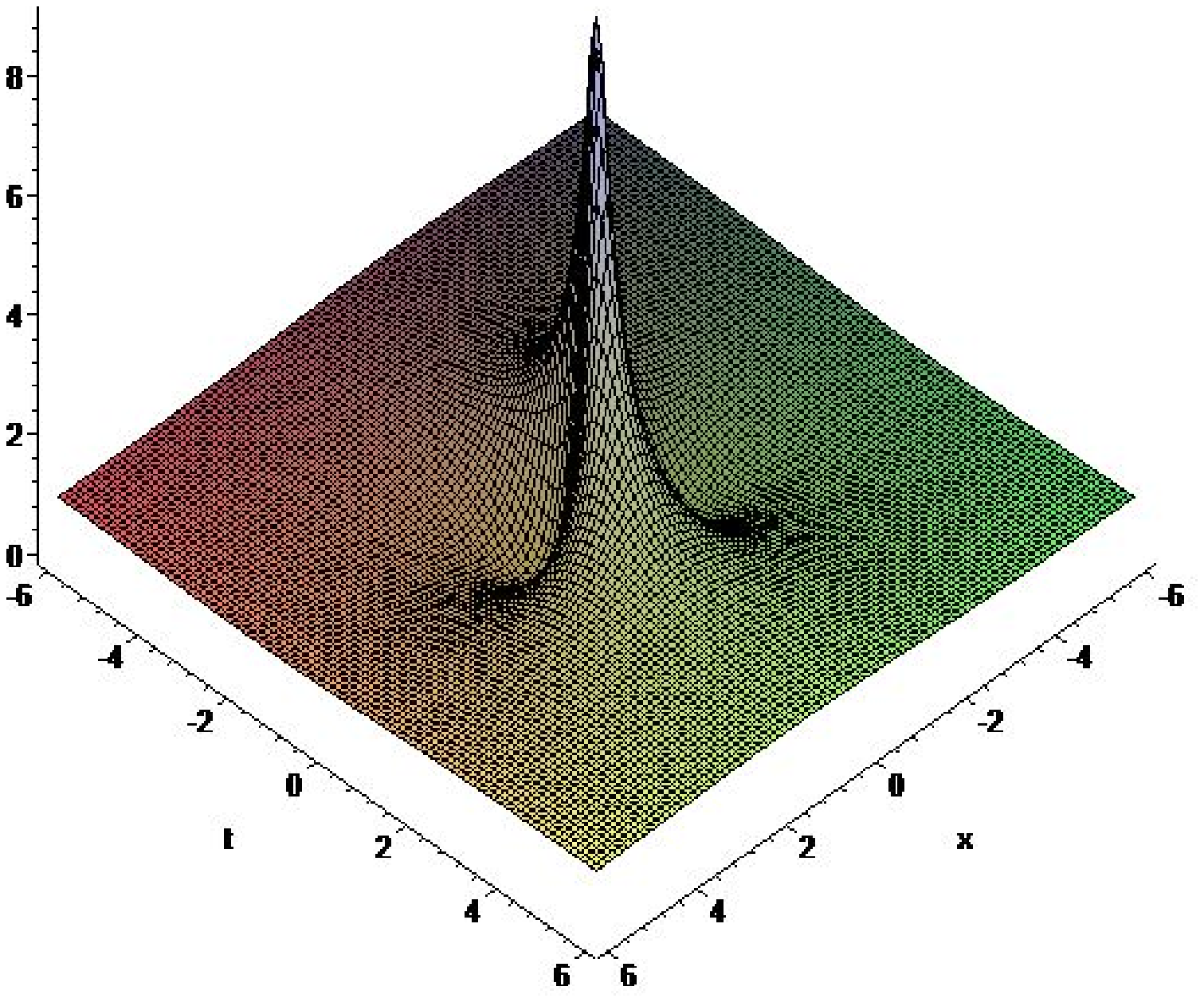}
\hskip 0.01cm
\raisebox{0.85in}{}\raisebox{-0.1cm}{\includegraphics[scale=0.23]{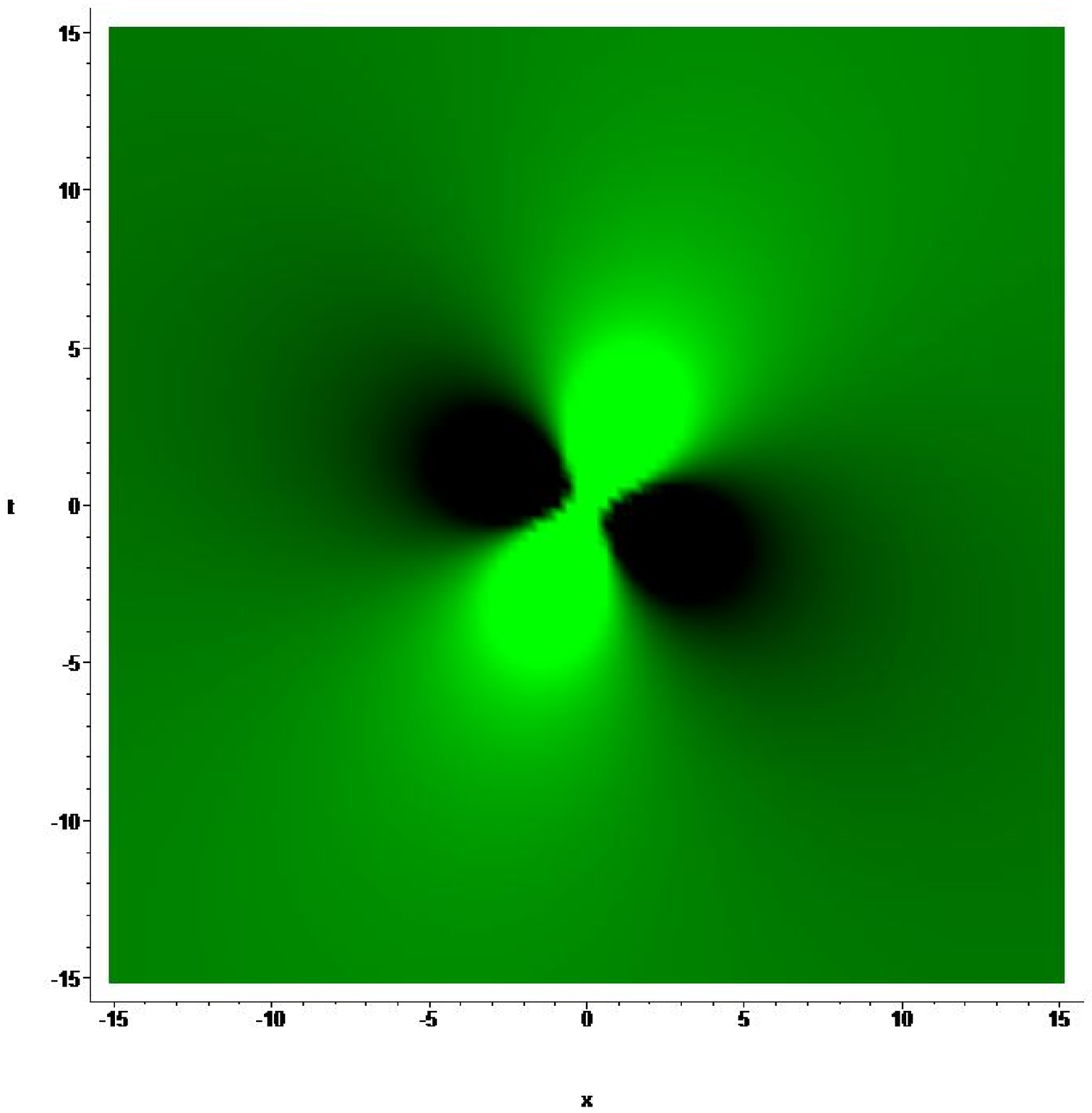}}
%\hskip 0.01cm
%\raisebox{0.85in}{($\eta$)}\raisebox{-0.1cm}{\includegraphics[scale=0.18]{V2-soliton-y}}
 \caption{One order rogue wave solution $|Q_{r}^{[1]}|^2$  of the Kundu-DNLS equation with
 $a=-2,c=1,\xi=1,\eta=1.$}\label{onerogue}
\end{figure}

In this section, we construct the rogue wave solution of Kundu-DNLS equation. This kind of solution only appears in some special region
of time and space and then drown into a fixed non-vanishing
plane. By making use of the Taylor expansion for the breather
solution, one order rogue wave solution of $Q_{r}^{[1]}$ for the Kundu-DNLS equation is
obtained
\begin{eqnarray}
&&Q_{r}^{[1]}=\frac{-v_{1}e^{-i(2x+t)}}{v_{2}},\nonumber\\
\end{eqnarray}
where
\begin{eqnarray*}
v_{1}&=&3+8x^{2}+8itx^{2}+8ixt^{2}+8xt-8t^{2}x^{2}-4t^{4}-4x^{4}+8ix^{3}-4ix+12it+8it^{3}-8t^{2},\\
v_{2}&=&-1+8it^{2}+4it+8itx^{2}-8it^{2}x-8tx-8t^{2}x^{2}-8ix^{3}-4t^{4}-4x^{4}-4ix.
\end{eqnarray*}
The picture of one order rogue wave solution of the Kundu-DNLS equation and its corresponding density graph are plotted in Fig.\ref{onerogue} .
Comparing  Fig.\ref{1soliton} and Fig.\ref{onerogue} can tell us the following important difference between one rogue wave solution and  one soliton solution.
Rogue wave solution is localized in both x and t direction which means it only has large amplitude with a non-vanishing boundary in a small domain of (x,t) plane.
  But a soliton solution is a travelling wave and has a vanishing boundary.%%%%%%%%%%%%%%%%%%%%%%%%%%%%%
%%%%%%%%%%%%%%%%%%%%%%%%%%%%%%%%%%%%%%%%%%%%%%%%%%%%%%%%%%%

When we take $a=-2,c = 1,\varepsilon=1,\eta=1$, the specific form of
the second rogue wave solution for the Kundu-DNLS equation takes the following form:
\begin{eqnarray}
&&Q_{r}^{[2]}=\frac{-v_{3}v_{4}e^{-i(2x+t)}}{v_{5}^{2}},\nonumber\\
\end{eqnarray}
where
\begin{eqnarray*}
v_{3}&=&-72xt+48x^{3}t-216x^{2}t^{2}+24x^{2}t^{4}+24x^{4}t^{2}+90x^{2}+666t^{2}-12x^{4}+180t^{4}+8t^{6}+8x^{6}\\
&&+48xt^{3}+9-48ix^{3}-48ix^{3}t^{2}+288ixt^{2}-54ix-24ixt^{4}+24it^{5}+24ix^{4}t+198it+336it^{3}\\&&+48ix^{2}t^{3}-24ix^{5},\\ v_{4}&=&198x^{2}-45-504xt+144x^{3}t+504x^{2}t^{2}+144xt^{3}+486t^{2}+60t^{4}+60x^{4}-24x^{2}t^{4}-8t^{6}\\
&&-24x^{4}t^{2}-8x^{6}-48ix^{3}+24ix^{5}+48ix^{3}t^{2}+24ixt^{4}-288ix^{4}t-576ixt^{2}+144ix^{2}t^{3}-90ix\\
&&-414it+72ix^{4}t+528it^{3}+72it^{5},\\
v_{5}&=&-48ix^{3}-48ix^{3}t^{2}+288ixt^{2}-54ix-24ixt^{4}+72xt-48x^{3}t+216x^{2}t^{2}-24x^{2}t^{4}-24ix^{5}\\
&&-90x^{2}-666t^{2}+24it^{5}+12x^{4}-180t^{4}-8t^{6}-8x^{6}-48xt^{3}+24ix^{4}t+198it+336it^{3}-9\\
&&+48ix^{2}t^{3}-24x^{4}t^{2}.
\end{eqnarray*}
The picture of second order rogue wave solution of the Kundu-DNLS equation and its corresponding density graph are plotted in Fig.6 .
%%%%%%%%%%%%%%%%%%%%%%%%%%%%%
\begin{figure}[h!]
\centering
\raisebox{0.85in}{}\includegraphics[scale=0.39]{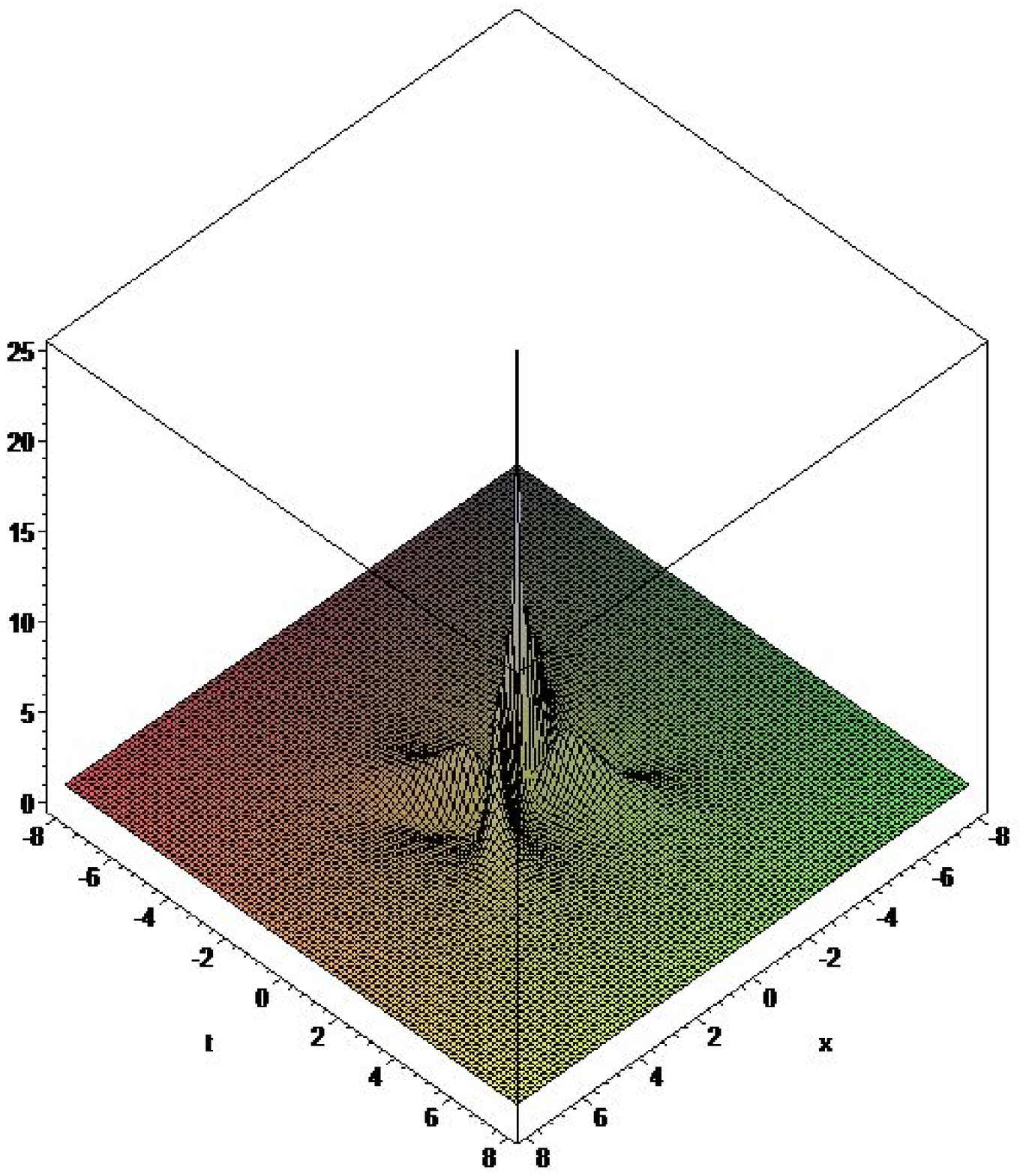}
\hskip 0.01cm
\raisebox{0.85in}{}\raisebox{-0.1cm}{\includegraphics[scale=0.23]{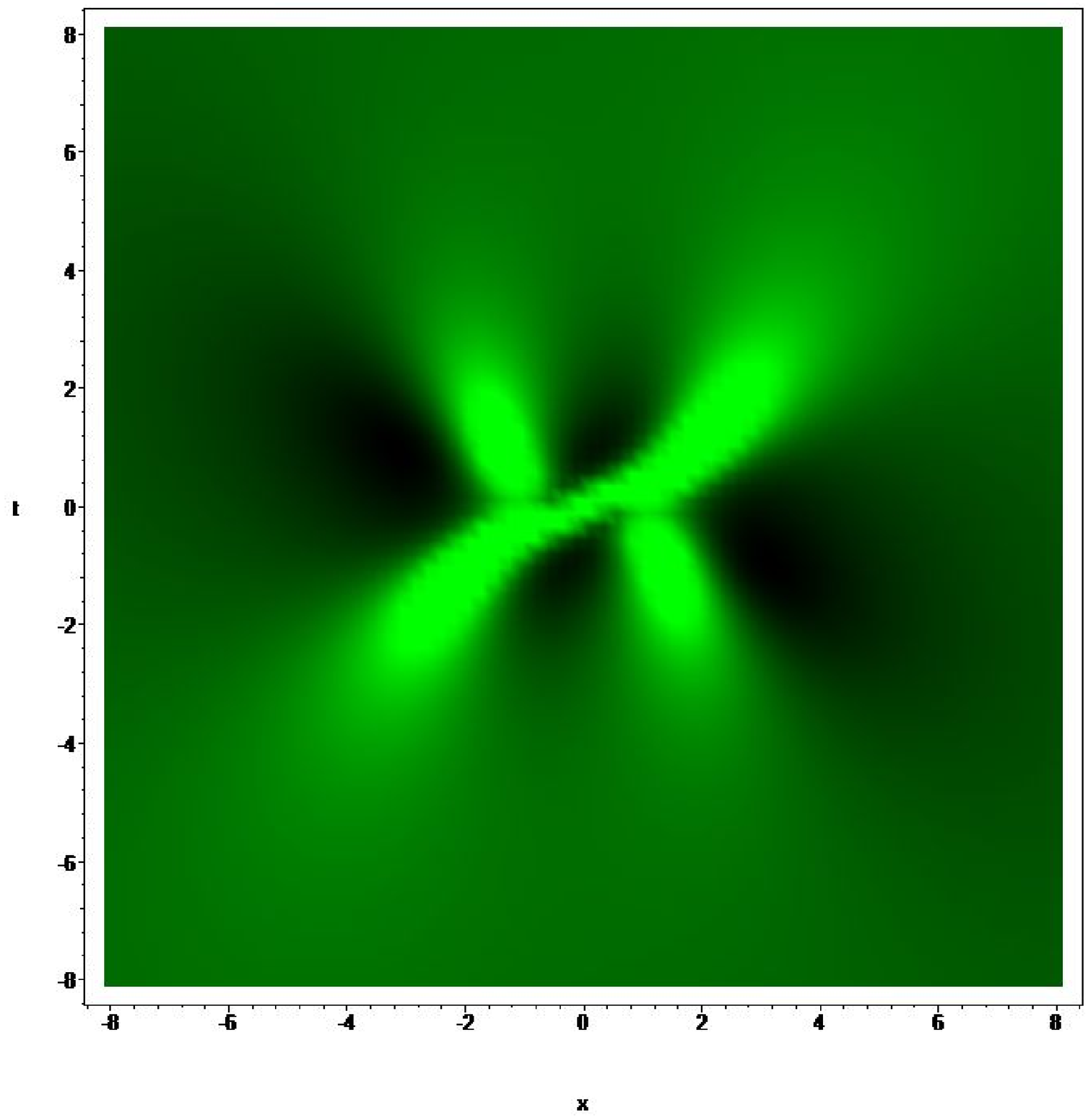}}
%\hskip 0.01cm
%\raisebox{0.85in}{($\eta$)}\raisebox{-0.1cm}{\includegraphics[scale=0.18]{V2-soliton-y}}
 \caption{The second rogue wave solution $|Q_{r}^{[2]}|^2$  of the Kundu-DNLS equation with
 $a=-2,c=1,\xi=1,\eta=1.$}\label{positon}
\end{figure}
%%%%%%%%%%%%%%%%%%%%%%%%%%%%%%%%%%%%%%%%%%%%%%%%%%%%%%%%%%%

Next, if we replace\eqref{eigenfunfornonzeroseed} with the following expression,
\begin{eqnarray}\notag
\left(\mbox{\hspace{-0.2cm}} \begin{array}{c}
 \phi_{k}(x,t,\lambda_{k})\\
 \varphi_{k}(x,t,\lambda_{k})\\
\end{array}\mbox{\hspace{-0.2cm}}\right)\mbox{\hspace{-0.2cm}}=\mbox{\hspace{-0.2cm}}\left(\mbox{\hspace{-0.2cm}}\begin{array}{c}
D_1f_1(x,t,\lambda_{k})[1,k]+D_2f_2(x,t,\lambda_{k})[1,k]+D_2f_1^{\ast}(x,t,{\lambda_{k}^{\ast})}[2,k]+D_1f_2^{\ast}(x,t,{\lambda_{k}^{\ast})}[2,k]\\
D_1f_1(x,t,\lambda_{k})[2,k]+D_2f_2(x,t,\lambda_{k})[2,k]+D_2f_1^{\ast}(x,t,{\lambda_{k}^{\ast})}[1,k]+D_1f_2^{\ast}(x,t,{\lambda_{k}^{\ast})}[1,k]\\
\end{array}\mbox{\hspace{-0.2cm}}\right)\\
\end{eqnarray}
where

\begin{equation*}\label{D3}
 \left\{
    \begin{split}
      D_1=&\exp\left(-{\rm i}s(S_0+S_1\epsilon+S_2\epsilon^{2})\right)\\
      D_2=&\exp\left({\rm i}s(S_0+S_1\epsilon+S_2\epsilon^{2})\right).
    \end{split}
    \right.
 \end{equation*}
We can split the second order rogue wave solution into triangle structure with the help of $S_1$. A particular structure is displayed
in Fig. 7. It is easy to see that three intensity humps appear at different times and space, and each intensity hump is roughly a first-order rogue wave.
\begin{figure}[h!]
\centering
\raisebox{0.85in}{}\includegraphics[scale=0.35]{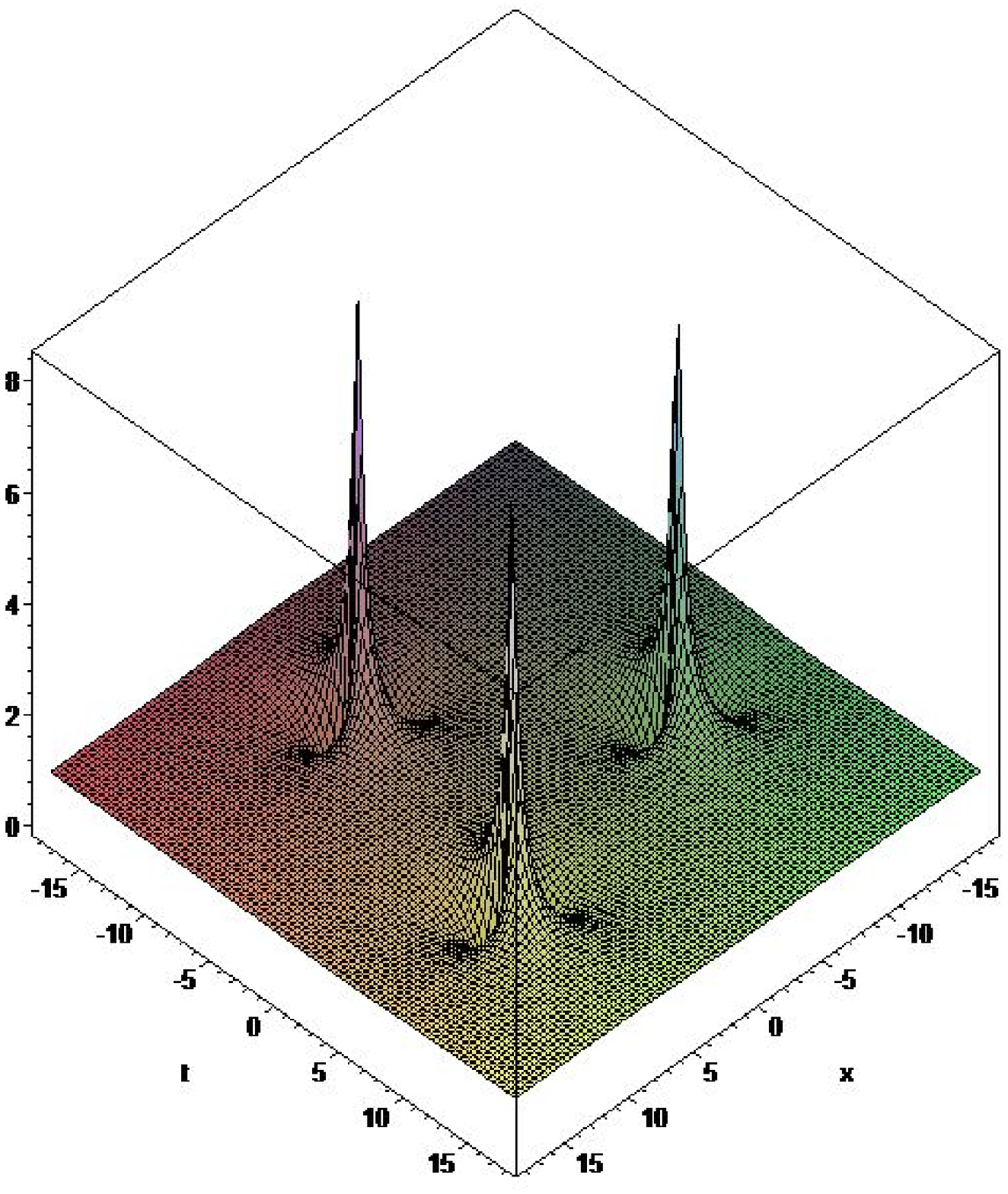}
\hskip 0.01cm
\raisebox{0.85in}{}\raisebox{-0.1cm}{\includegraphics[scale=0.27]{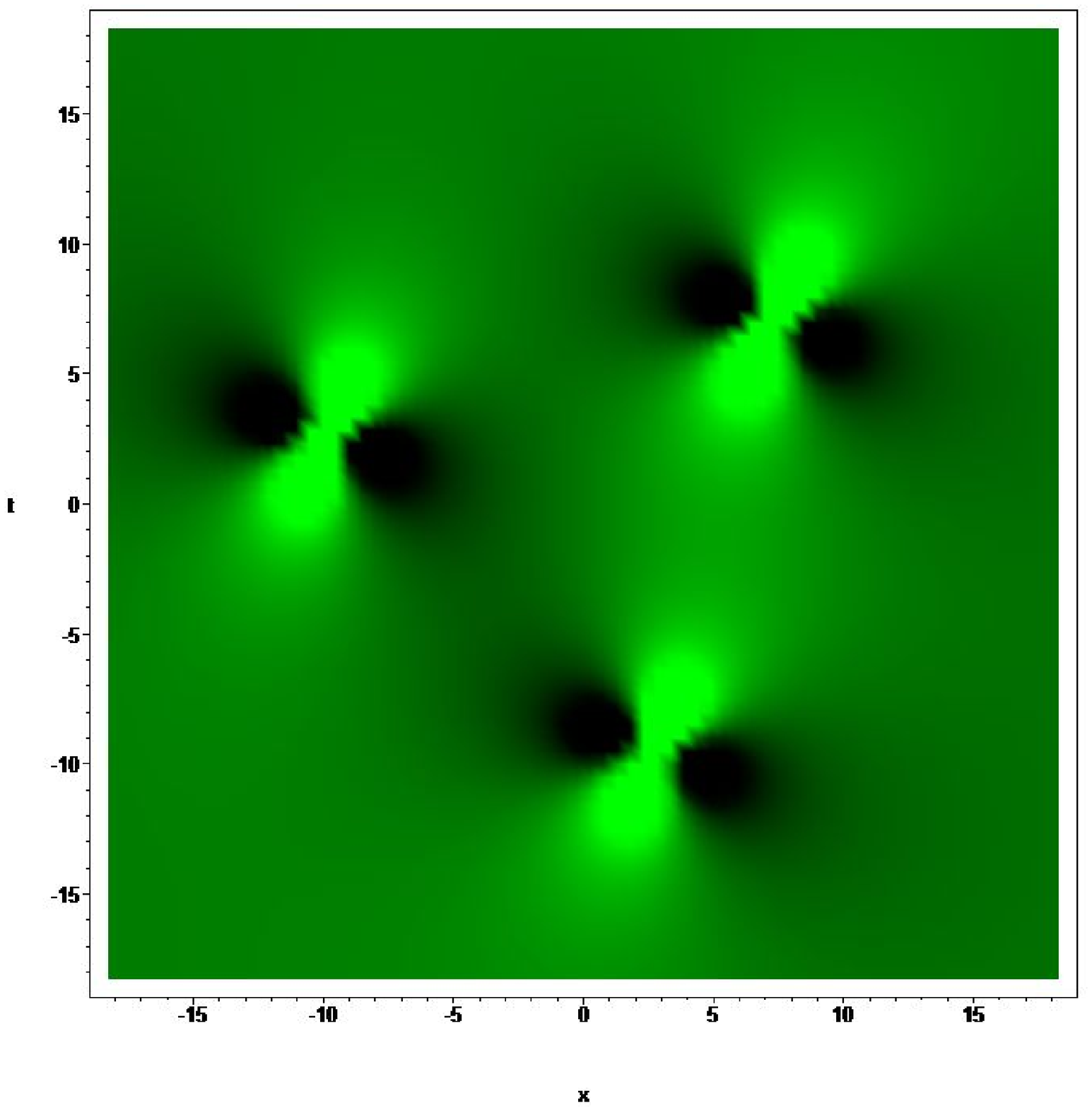}}
%\hskip 0.01cm
%\raisebox{0.85in}{($\eta$)}\raisebox{-0.1cm}{\includegraphics[scale=0.18]{V2-soliton-y}}
 \caption{The second rogue wave solution $|Q_{r}^{[2]}|^2$  of the Kundu-DNLS equation with
 $S_0=0, S_1=500, S_2=0.$}\label{positon}
\end{figure}

%%%%%%%%%%%%%%%%%%%
Next, we examine third-order rogue waves. In this case,  Form the figures, We can get third-order rogue wave solution with the help of $a=-2, c=1, \alpha=1$. A particular structure is displayed
in Fig. 8.
\begin{figure}[h!]
\centering
\raisebox{0.85in}{}\includegraphics[scale=0.39]{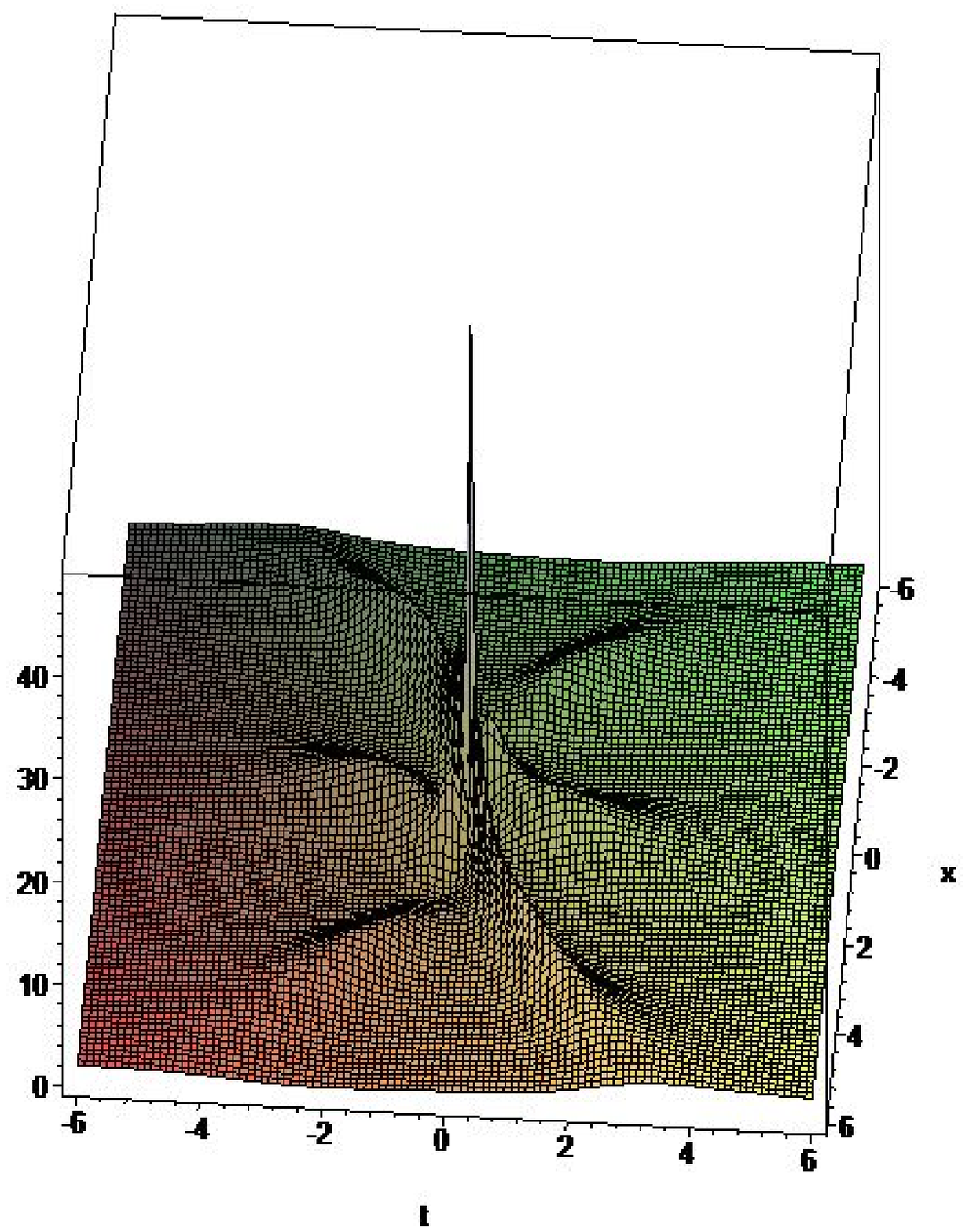}
\hskip 0.01cm
\raisebox{0.85in}{}\raisebox{-0.1cm}{\includegraphics[scale=0.27]{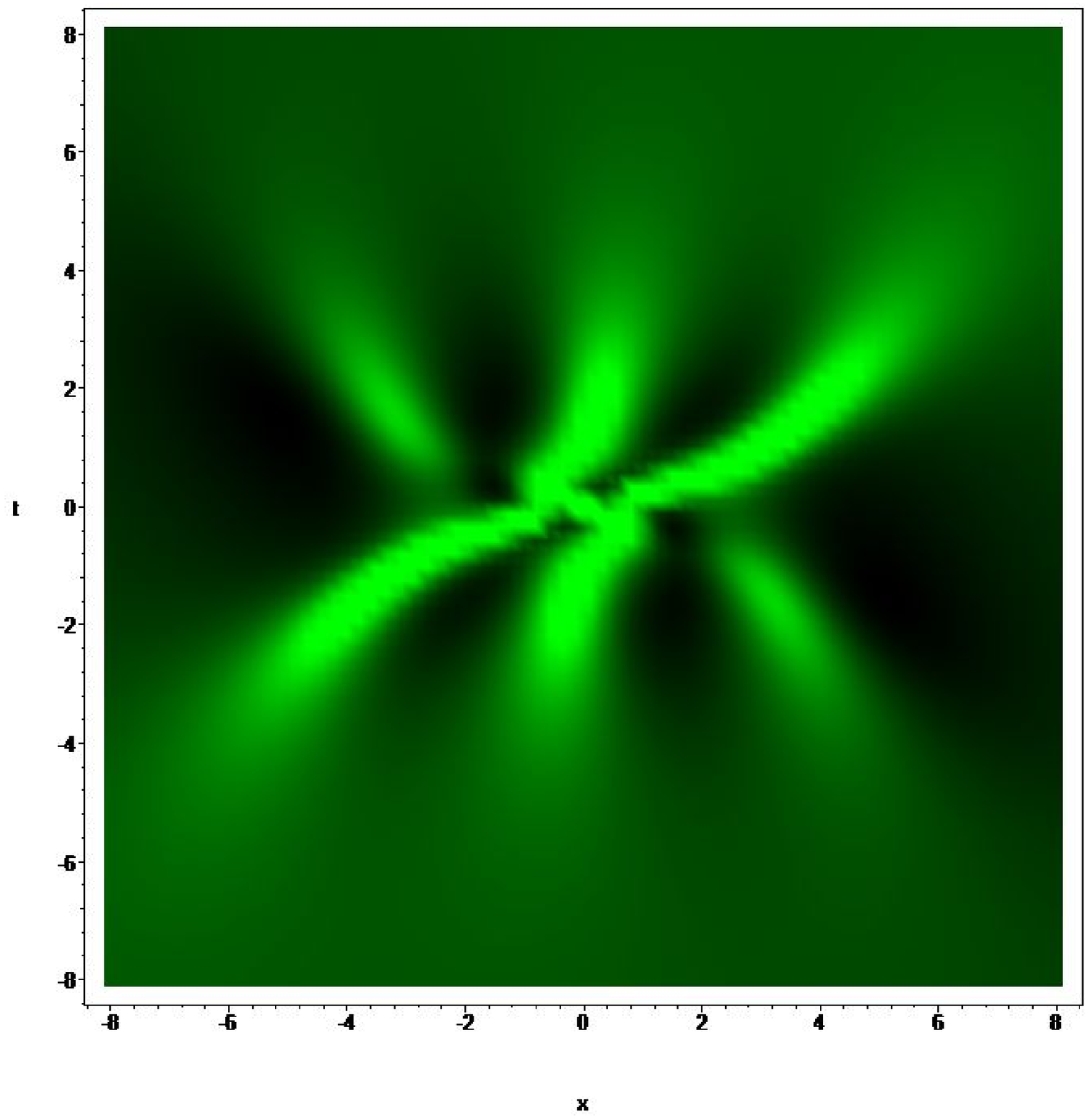}}
%\hskip 0.01cm
%\raisebox{0.85in}{($\eta$)}\raisebox{-0.1cm}{\includegraphics[scale=0.18]{V2-soliton-y}}
 \caption{The third rogue wave solution $|Q_{r}^{[3]}|^2$  of the Kundu-DNLS equation with
 $a=-2, c=1,\alpha=1, \varepsilon=0.8, \eta=0.8.$}\label{positon}
\end{figure}
%%%%%%%%%%%%%%%%%%%%%%%%%%%

%%%%%%%%%%%%%%%%%%%
We can split the third order rogue wave solution into triangle structure with the help of $S_1$. A particular structure is displayed
in Fig. 9. The third-order rogue wave is seen to possess a regular triangle spatial symmetry structure.
\begin{figure}[h!]
\centering
\raisebox{0.85in}{}\includegraphics[scale=0.35]{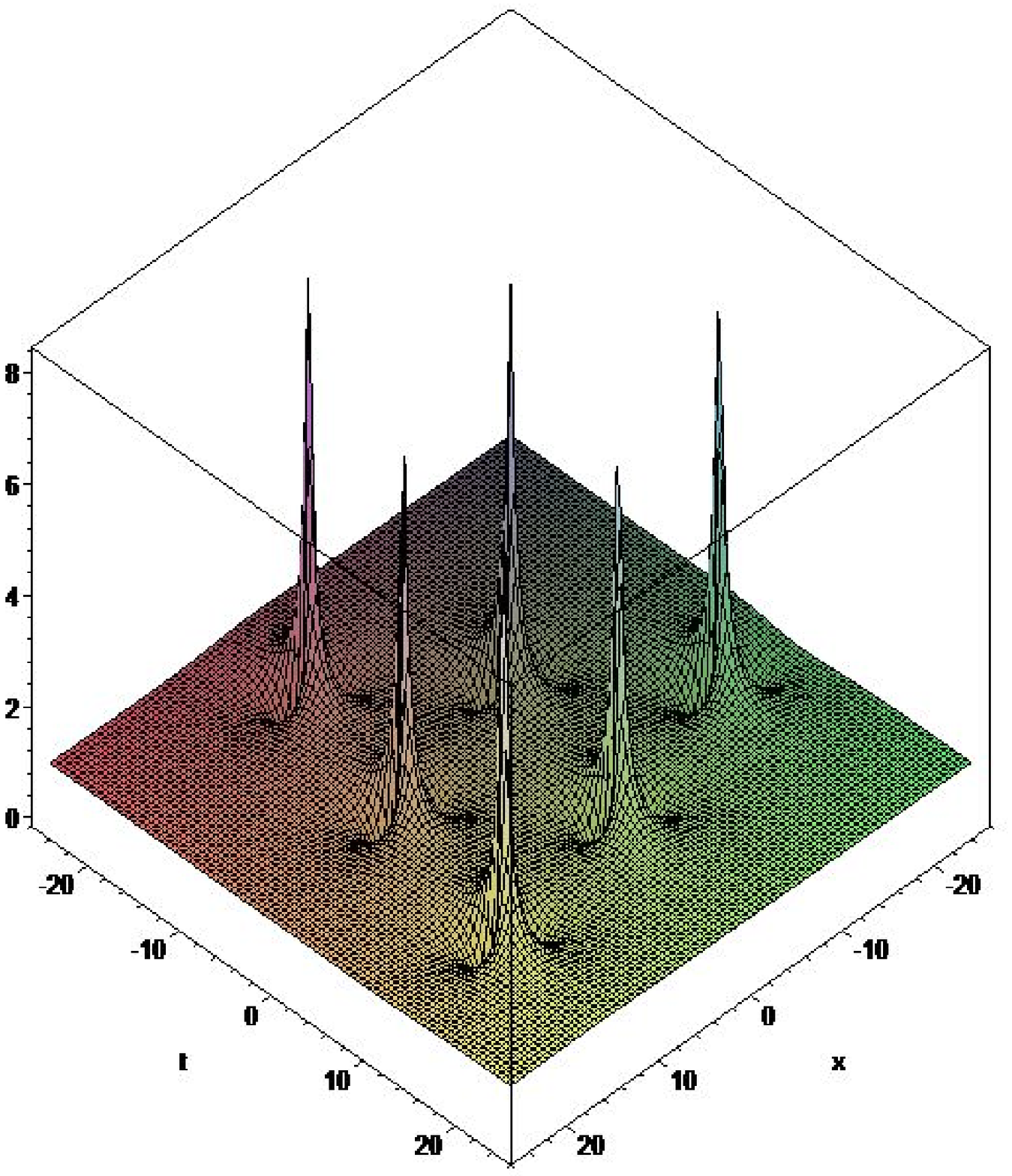}
\hskip 0.01cm
\raisebox{0.85in}{}\raisebox{-0.1cm}{\includegraphics[scale=0.25]{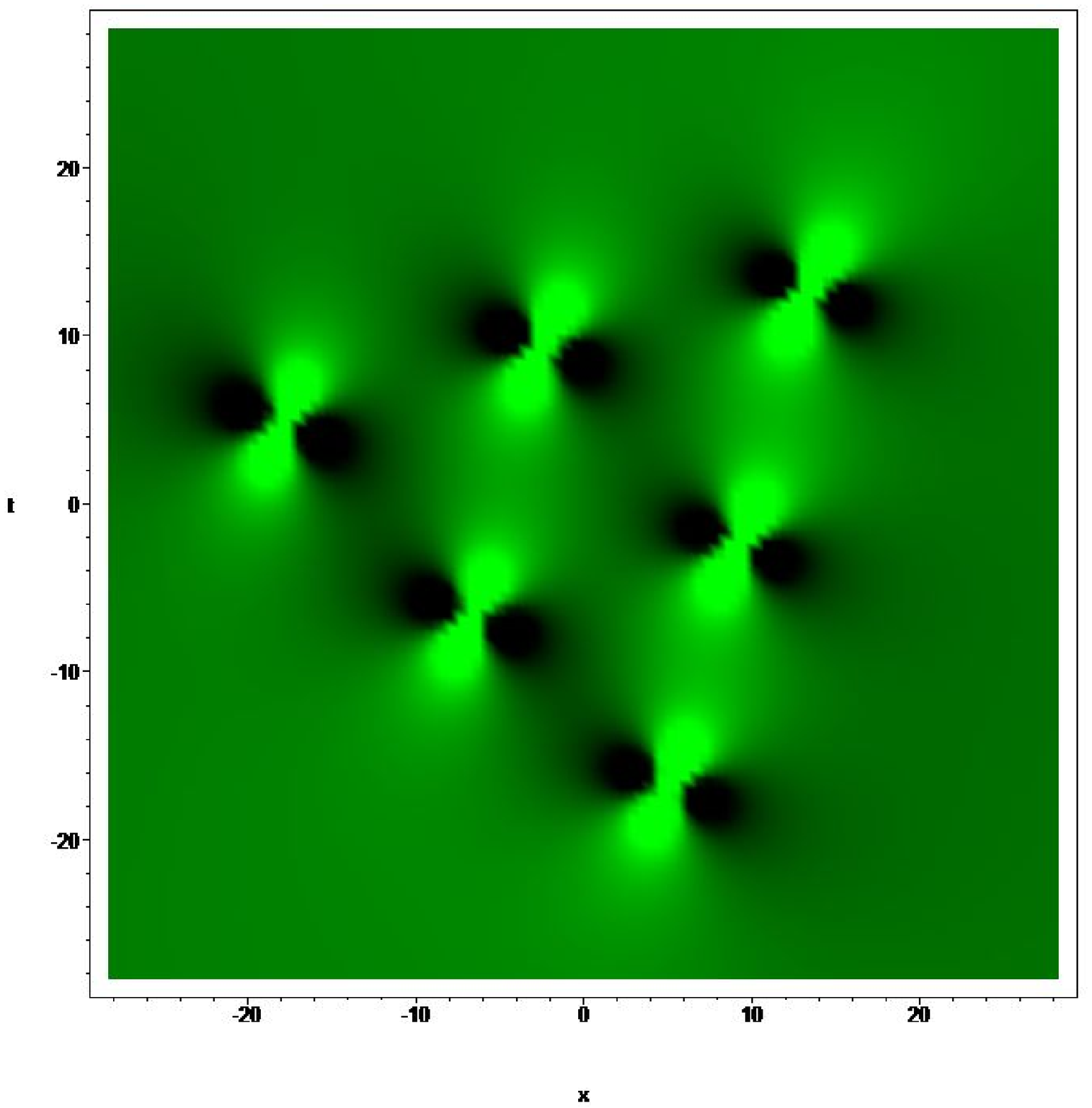}}
%\hskip 0.01cm
%\raisebox{0.85in}{($\eta$)}\raisebox{-0.1cm}{\includegraphics[scale=0.18]{V2-soliton-y}}
 \caption{The third rogue wave solution $|Q_{r}^{[3]}|^2$  of the Kundu-DNLS equation with
 $a=-2, c=1, \alpha=1, S_0=0, S_1=500, S_2=0.$}\label{positon}
\end{figure}
%%%%%%%%%%%%%%%%%%%%%%%%%%%

We can split the third order rogue wave solution into pentagon structure with the help of $S_2$. A particular structure is displayed
in Fig. 10. The third-order rogue wave exhibits a regular pentagon spatial symmetry structure.
\begin{figure}[h!]
\centering
\raisebox{0.85in}{}\includegraphics[scale=0.39]{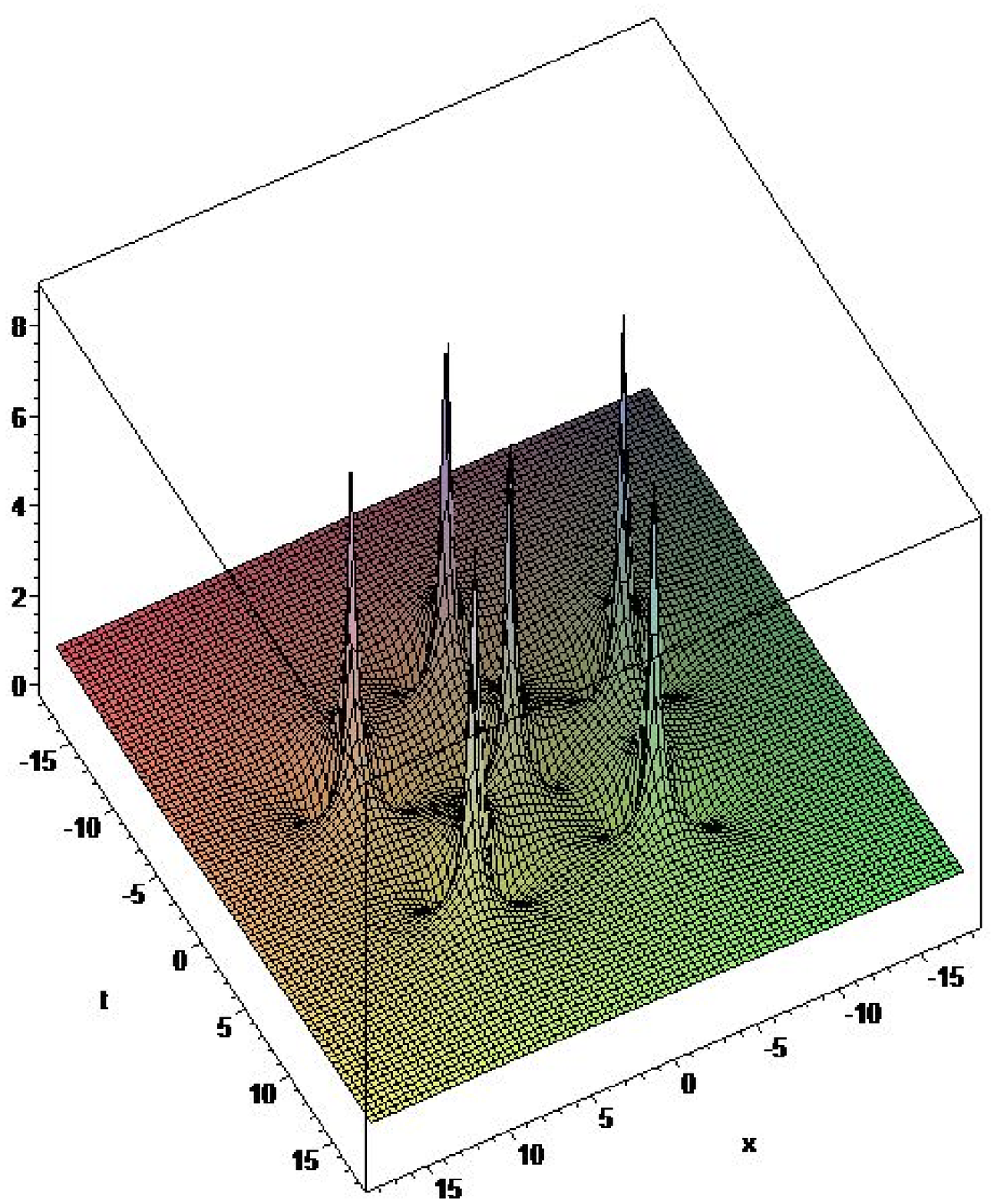}
\hskip 0.01cm
\raisebox{0.85in}{}\raisebox{-0.1cm}{\includegraphics[scale=0.27]{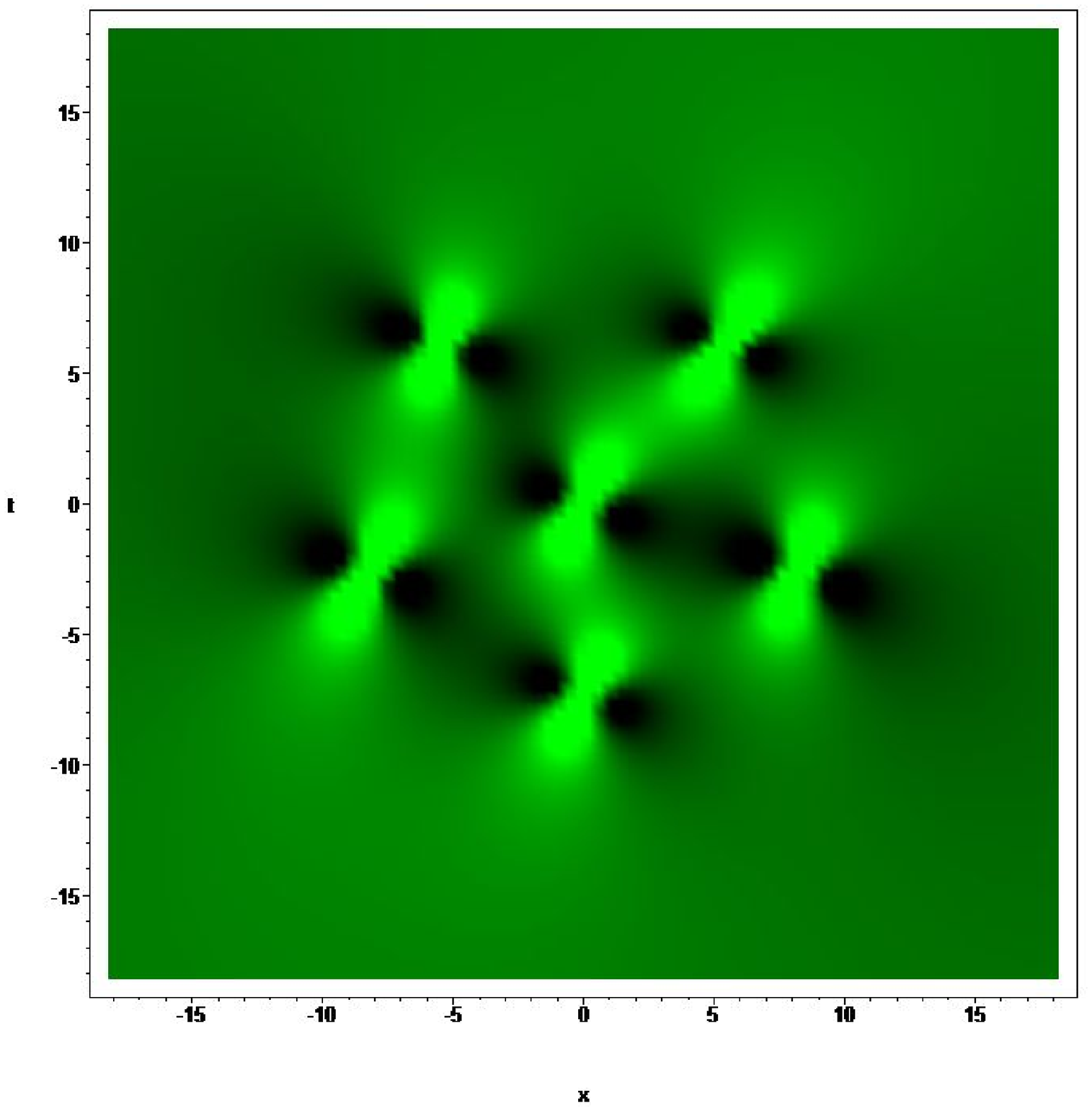}}
%\hskip 0.01cm
%\raisebox{0.85in}{($\eta$)}\raisebox{-0.1cm}{\includegraphics[scale=0.18]{V2-soliton-y}}
 \caption{The third rogue wave solution $|Q_{r}^{[3]}|^2$  of the Kundu-DNLS equation with
 $a=-2,c=1,\alpha=1,S_0=0,S_1=0,S_2=1000.$}\label{positon}
\end{figure}
%%%%%%%%%%%%%%%%%%%%%%%%%%%

From above graphs of rogue wave solutions of the Kundu-DNLS equation, we can find some twisted effect of modified terms of the Kundu-DNLS equation which is different from DNLS equation\cite{SWX}.

%%%%%%%%%%%%%%%%%%%%%%%%%%%%%%%%%%%%%%%%%%%%%%%%%%%%%%%%%%%%%%
\section{Conclusions}
In this paper, we construct the Darboux transformation for the Kundu-DNLS equation. And the determinant representations of the new solution $Q$ of the Kundu-DNLS equation
are given. Moreover, by making use of the Darboux transformation, we derive several types of solutions for Kundu-DNLS equation. There solutions include the soliton solutions, positon solution and breather solution, 1-order rogue wave, 2-order rogue wave and 3-order rogue wave. Particularly, these rogue wave solutions possess several free parameters. With the help of these parameters, these rogue waves constitute some patterns,  such as fundamental pattern,  triangular pattern, circular pattern. On the other hand, we can also derive the higher order rogue wave solutions for Kundu-DNLS equation by making use of the Darboux transformation. The application of high-order rogue in physics will be one interesting subject.

{\bf Acknowledgments} {\noindent \small This work is supported by
the NSF of China under Grant No.11271210 and K. C. Wong Magna Fund in
Ningbo University. Jingsong He is also supported by Natural Science Foundation of Ningbo
under Grant No. 2011A610179. Chuanzhong Li is supported by the National
 Natural Science Foundation of China under Grant No.11201251,
 the Natural Science Foundation of Zhejiang
Province under Grant No. LY12A01007. We thank Prof. Yishen Li (USTC, Hefei,
China) for his long time support and useful suggestions. }

%%%%%%%%%%%%%%%%%%%%%%%%%%%%%%%%%%%%%%%%%%%%%%

%---------------------------------------------------------------------------------------

\end{document}